\documentclass[11pt,a4paper]{iopart}
\usepackage{epsfig}
\begin{document}
\title{A Lovelock black hole bestiary}

\author{Xi\'an O. Camanho$^\dag$ and Jos\'e D. Edelstein$^{\dag\,\ddag}$}

\address{$\dag$ Department of Particle Physics and IGFAE, University of Santiago de Compostela, E-15782 Santiago de Compostela, Spain}\vskip1mm
\address{$\ddag$ Centro de Estudios Cient\'\i ficos, Valdivia, Chile}
\eads{\mailto{xian.otero@rai.usc.es}, \mailto{jose.edelstein@usc.es}}

\begin{abstract}
We revisit the study of (A)dS black holes in Lovelock theories. We present a new tool that allows to attack this problem in full generality. In analyzing maximally symmetric Lovelock black holes with non-planar horizon topologies many distinctive and interesting features are observed. Among them, the existence of maximally symmetric vacua do not supporting black holes in vast regions of the space of gravitational couplings, multi-horizon black holes, and branches of solutions that suggest the existence of a rich diagram of phase transitions. The appearance of naked singularities seems unavoidable in some cases, raising the question about the fate of the cosmic censorship conjecture in these theories. There is a preferred branch of solutions for planar black holes, as well as non-planar black holes with high enough mass or temperature. Our study clarifies the role of all branches of solutions, including asymptotically dS black holes, and whether they should be considered when studying these theories in the context of AdS/CFT.
\end{abstract}
\submitto{\CQG}

\section{Introduction}

Lovelock gravities are the higher dimensional cousins of general relativity. If one assumes that the spacetime dimensionality is larger than four, the lagrangian that embodies Einstein's paradigm {\it gravity equals geometry}, while providing second order Euler-Lagrange equations for the metric tensor, has been constructed by David Lovelock some forty years ago \cite{Lovelock1971}. The simplest Lovelock lagrangian is the well-known Gauss-Bonnet term, that embodies a non-trivial dynamics for the gravitational field in five (or higher) dimensional theories. It has been explored to a large extent, possibly due to the appealing place that five-dimensional gravity occupies since Kaluza and Klein's groundbreaking papers, and also as a consequence of its appearance in string theories at low energies \cite{Zwiebach1985}.

Since their inception, a steady attention has been devoted to scrutinize the main properties of Lovelock theories of gravity, their vacuum structure, induced cosmologies, hamiltonian formalism, dimensional reduction, wormhole configurations and, most importantly, their black hole solutions, including their formation, stability and thermodynamics. In spite of the abundant literature on the subject, most articles deal with particular cases of the general Lovelock formalism due to the intricacy endowed by the increasing number of coupling constants: there are $[\frac{d-3}{2}]$ dimensionful quantities (alongside the Newton and cosmological constants) in a $d$-dimensional theory. For this reason, many investigations on black hole solutions of Lovelock gravities are restricted to one-parameter (zero measure) subspaces in the space of couplings. It is the aim of this article to tackle the existence and main features of Lovelock black holes for arbitrary values of the full set of gravitational couplings. 

Despite its debatable phenomenological interest, Lovelock gravities provide an interesting framework from a theoretical point of view for several reasons. As higher dimensional members of Einstein's general relativity family, they allow to explore several conceptual issues of gravity in depth in a broader setup. Among these, we can include features of black holes such as their existence and uniqueness theorems, their thermodynamics, the definition of their mass and entropy, etc. Lovelock theories are perfect toy models to contrast our ideas about gravity. We shall also recall at this point that much efforts have been devoted in the last quarter of a century in high energy physics dealing with scenarios involving higher dimensional gravity, and it is fair to say that at present it is unclear if gravity is a truly four-dimensional interaction.

A final piece of motivation comes from the theoretical framework proposed by Juan Maldacena \cite{Maldacena1998}. The AdS/CFT correspondence establishes a holographic identification between conformal field theories and quantum gravities in higher dimensional AdS spaces. Besides its original maximally supersymmetric formulation, the correspondence seems robust enough to survive its generalization to less supersymmetric scenarios \cite{Klebanov1998}, and even non-supersymmetric \cite{Polyakov1999}, as well as non-stringy realizations \cite{Strominger:1997eq} (see also the seminal paper \cite{Brown1986b}). In particular, even if some caution remarks should be quoted at this point, the AdS/CFT correspondence seems to apply in higher dimensions too.

We know very little about non-trivial conformal field theories in higher dimensions (see \cite{Witten2007c} for a recent discussion). Still, we can formally compute $2$-point and $3$-point stress-energy correlation functions and use positivity of the energy to show that the central charges are constrained to certain values \cite{Boer2009,Camanho2010,Buchel2010a,Camanho2010a,Boer2009a} (this was originally discussed in the case of four-dimensional CFTs in \cite{Hofman2008,Buchel2009a,Hofman2009}). Strikingly enough, these restrictions tantamount to analogous conditions on the gravitational coupling constants due to potential superluminal states propagating at the boundary of AdS that correspond to highly energetic gravitons exploring the bulk \cite{Brigante2008}. These superluminal states have been proven to exist in the CFT side when positivity of the energy constraints are violated \cite{Kulaxizi2011}.

Applications of AdS/CFT towards the understanding of the hydrodynamics of CFT plasmas in arbitrary dimensions demand a proper understanding of Lovelock black holes in AdS. This provides the final bit of motivation to pursue the present investigation. Regardless of the phenomenological dimensionality required by these applications, it is customarily the case that pushing some ideas to their extremes, besides verifying their robustness, allows to discover novel features that are hidden in the somehow simpler original formulation (see, for instance, \cite{Paulos2011} for a beautiful recent example of this statement).

The paper is organized as follows. We give a brief introduction on the main features of Lovelock gravity in section 2. Section 3 is devoted to present our proposal to deal with generic black holes in Lovelock theory. Finding an analytic black hole solution requires to explicitly solve a polynomial equation and we are certainly restricted by the implications of Galois theory; meaningly, quartic is the highest order polynomial equation that can be generically solved by radicals (Abel-Ruffini theorem). However, an implicit but exact solution can be found, and we develop some tools to extract all relevant information, mainly their horizon structure and thermodynamics.

We perform a detailed classification of all possible black hole solutions in section 4, including the case of asymptotically dS solutions, and all possible horizon topologies within maximally symmetric configurations. Section 5 is devoted to study the specific heat of the whole family of black holes and we discuss their local thermodynamic stability. We find that Hawking-Page-like phase transitions should generically be present, a discussion that is part of section 6. Our analysis leads to some puzzles about the cosmic censorship conjecture in Lovelock theories. We end by discussing our results and commenting on some avenues for future research in section 7.

~

\noindent
{\it Note added}: While finishing the writing up of this article we became aware of related research being pursued by H. Maeda, S. Willison and S. Ray \cite{Maeda2011}.

\section{Lovelock gravity}

Lovelock gravity is the most general second order gravity theory in higher-dimensional spacetimes, and it is free of ghosts when expanding about flat space \cite{Lovelock1971,Zumino1986}. The bulk action can be written in terms of differential forms as
\begin{equation}
\mathcal{I} =\frac{1}{16\pi G_N (d-3)!}\, \sum_{k=0}^{K} {\frac{c_k}{d-2k}} \mathop\int \mathcal{R}^{(k)} ~,
\label{LLaction}
\end{equation}
$G_N$ being the Newton constant in $d$ spacetime dimensions. $c_k$ is a set of couplings with length dimensions $L^{2(k-1)}$, $L$ being a length scale related to the cosmological constant, while $K$ is a positive integer,
\begin{equation}
K\leq \left[\frac{d-1}{2}\right] ~,
\label{maximalK}
\end{equation}
labeling the highest non-vanishing coefficient, {\it i.e.}, $c_{k>K} = 0$. $\mathcal{R}^{(k)}$ is the exterior product of $k$ curvature 2-forms, $
R_{~b}^a = d\,\omega_{~b}^a + \omega_{~c}^a \wedge \omega_{~b}^c =  \frac{1}{2} R_{~b\mu\nu}^{a}\; dx^{\mu} \wedge dx^{\nu}$, $\omega_{~b}^a$ being the 1-form spin connection, with the required number of vielbein, $e^a$, to construct a $d$-form,
\begin{equation}
\mathcal{R}^{(k)} = \epsilon_{f_1 \cdots f_{d}}\; R^{f_1 f_2} \wedge \cdots \wedge R^{f_{2k-1} f_{2k}} \wedge e^{f_{2k+1}} \wedge \ldots \wedge e^{f_d} ~.
\end{equation}
The zeroth and first term in (\ref{LLaction}) correspond, respectively, to the cosmological term and the Einstein-Hilbert action. These are particular cases of the Lovelock theory. It is fairly easy to see that $c_0 = L^{-2}$ and $c_1 = 1$ correspond to the usual normalization of these terms, the cosmological constant having the customary value $2 \Lambda = - (d-1)(d-2)/L^2$. Either a negative ($c_0 = -L^{-2}$) or a vanishing ($c_0 = 0$) cosmological constant can be easily incorporated into our approach.

If we consider this action in the first order formalism, we have two equations of motion, for the connection 1-form and for the vierbein. If we vary the action with respect to the connection the resulting equation is proportional to the torsion, $T^a = d\,e^a + \omega_{~b}^a \wedge e^b$, so we can safely set it to zero as usual, allowing us to compare our results with those coming from the usual tensorial formalism based on the metric.

The second equation of motion is obtained by varying the action with respect to the vierbein. It can be cast into the form
\begin{equation}
\mathcal{E}_a \equiv \epsilon_{a f_1 \cdots f_{d-1}}\; \mathcal{F}_{(1)}^{f_1 f_2} \wedge \cdots \wedge \mathcal{F}_{(K)}^{f_{2K-1} f_{2K}} \wedge e^{f_{2K+1}} \wedge \ldots \wedge e^{f_{d-1}} = 0 ~,
\label{eqlambda}
\end{equation}
where $\mathcal{F}_{(i)}^{a b} \equiv R^{a b} - \Lambda_i\, e^a \wedge e^b$, which makes manifest that, in principle, this theory admits $K$ constant curvature {\it vacuum} solutions,
\begin{equation}
\mathcal{F}_{(i)}^{a b} = R^{a b} - \Lambda_i\, e^a \wedge e^b = 0 ~.
\end{equation}
Inserting $R^{ab}=\Lambda\,e^a \wedge e^b$ in (\ref{eqlambda}), one finds that the $K$ different cosmological constants are the solutions of the $K$-th order polynomial
\begin{equation}
\Upsilon[\Lambda] \equiv \sum_{k=0}^{K} c_k\, \Lambda^k = c_K \prod_{i=1}^K \left( \Lambda - \Lambda_i\right) =0 ~,
\label{cc-algebraic}
\end{equation} 
each one corresponding to a different {\it vacuum}. The theory will have degenerate behavior whenever two or more effective cosmological constants coincide. This is captured by the discriminant,
\begin{equation}
\Delta = \prod_{i < j}^K (\Lambda_i - \Lambda_j)^2 ~,
\label{discriminant}
\end{equation}
whose vanishment leads to a certain locus of the parameter space corresponding to the coupling constants of Lovelock theory. Curiously enough, most of the studies in the context of Lovelock theory have been performed within this degenerate locus. This article aims at making the complementary effort of digging into the non-degenerate case $\Upsilon[\Lambda_k]=0$, $\Upsilon'[\Lambda_k]\neq 0$, where $\Lambda_k$ is the {\it vacuum} under consideration. We will eventually see that, among the branches of solutions of (\ref{cc-algebraic}), only one would end up being physically relevant, at least for the high mass or high temperature regime\footnote{Both regimes are not equivalent and in principle one may have several masses corresponding to different branches in the high temperature regime. All such branches are perturbatively unstable in that limit except the one having infinite mass for infinite temperature \cite{Camanho2012}.}, say $\Lambda=\Lambda_\star$. Degeneracies that do not involve $\Lambda_\star$ are harmless, our analysis being thus valid for the whole parameter space, except the zero measure set $\Upsilon[\Lambda_\star]=\Upsilon'[\Lambda_\star]=0$.

\section{Lovelock black holes, a novel approach}

It has been shown in \cite{Aros2001} that Lovelock theories admit asymptotical (A)dS solutions with non-trivial horizon topologies. We can consider for instance solutions with a planar or hyperbolic symmetry as a straightforward generalization of the usual spherically symmetric ansatz,
\begin{equation}
ds^2 = - A(t,r)\, dt^2 + \frac{dr^2}{B(t,r)} + \frac{r^2}{L^2}\, d\Sigma_{d-2,\sigma}^2 ~,
\end{equation}
where
\begin{equation}
d\Sigma_{d-2,\sigma}^2 = \frac{d\rho^2}{1 - \sigma \rho^2/L^2}+\rho^2 d\Omega^2_{d-3} ~,
\end{equation}
is the metric of a $(d-2)$-dimensional manifold of negative, zero or positive constant curvature ($\sigma =-1, 0, 1$ parameterizing the different horizon topologies), and $d\Omega^2_{d-3}$ is the metric of the unit $(d-3)$-sphere. This does not imply that the horizon is just spherical or non-compact. By means of the Killing-Hopf theorem \cite{Wolf2011}, any complete connected Riemannian manifold of Euclidean signature and constant curvature $\sigma$ can be written as a quotient space $\Sigma_{d-2,\sigma}/\Gamma$, where $\Gamma$ is a discrete subgroup of the isometry group of $\Sigma_{d-2,\sigma}$. Thus, even in (what we shall call) the {\it spherical} case, we have non-spherical possibilities; for example, one may take the horizon to be a lens space. Besides, planar or hyperbolic horizons can be made compact in this way.

It has been proven in \cite{Zegers2005} that these black holes admit a version of Birkhoff's theorem, in such a way that in addition to the $SO(d-1)$, $E_{d-2}$ or $SO(1,d-2)$ isometry groups, these spacetimes admit  an extra timelike killing vector (for $A,B>0$). This means that these solutions of the field equations are locally isometric to their corresponding static counterparts, which can be found by means of the ansatz
\begin{equation}
ds^2 = - f(r)\, dt^2 + \frac{dr^2}{f(r)} + \frac{r^2}{L^2}\, d\Sigma_{d-2,\sigma}^2 ~.
\label{bhansatz}
\end{equation}
There are extra solutions with different functions in the timelike and radial direction but they are just valid for degenerate values of the cosmological constant \cite{Charmousis2002a}. In that case, the most general solution is
\begin{equation}
ds^2 = - f(r)\, dt^2 + \frac{dr^2}{(-\Lambda\, r^2)} + \frac{r^2}{L^2}\, d\Sigma_{d-2,\sigma}^2 ~,
\label{lifshitzbh}
\end{equation}
for any function $f(r)$. This allows in particular Lifshitz-like solutions $f(r)\sim r^{2z}$ for any value of the critical exponent $z$.

These black hole solutions are all three asymptotic to a maximally symmetric space. Thus, when considering the same curvature for all of them they are locally equivalent, but globally different. They are often referred to as topological black holes for this reason. Using the natural frame,
\begin{equation}
e^0 = \sqrt{f(r)}\, dt ~, \qquad e^1 = \frac{1}{\sqrt{f(r)}}\, dr ~, \qquad e^a = \frac{r}{L}\, \tilde e^a ~,
\label{vierbh}
\end{equation}
where $a = 2, \ldots, d-1$, and $\tilde R^{ab} = \sigma\, \tilde e^a \wedge \tilde e^b$. The Riemann 2-form reads
\begin{eqnarray}
& & R^{01} = - \frac12\, f''(r)\; e^0 \wedge e^1 ~, \quad\qquad R^{0a} = - \frac{f'(r)}{2 r}\; e^0 \wedge e^a ~, \nonumber \\ [1em]
& & R^{1a} = - \frac{f'(r)}{2 r}\; e^1 \wedge e^a ~, \qquad\qquad R^{ab} = - \frac{f(r) - \sigma}{r^2}\; e^a \wedge e^b ~.
\label{riemannbh}
\end{eqnarray}
If we insert these expressions into the equations of motion, we get
\begin{equation}
\left[ \frac{d~}{d\log r} + (d-1) \right]\, \left( \sum_{k=0}^{K} c_k\, g^k \right) = 0 ~,
\end{equation}
where $g = (\sigma- f)/r^2$. This can be readily solved as
\begin{equation}
\Upsilon[g]=\sum_{k=0}^{K} c_k\,  g^k = \frac{\kappa}{r^{d-1}} ~,
\label{eqg}
\end{equation} 
where $\kappa$ is an integration constant related to the mass of the spacetime \cite{Kastor2010,Kastor2011},
\begin{equation}
M=\frac{(d-2)V_{d-2}}{16\pi G_N}\, \kappa ~,
\label{mmass}
\end{equation} 
$V_{d-2}$ being the volume of the unit $(d-2)$-dimensional horizon. This can also be understood as follows. If there is actually a mass source for the gravitational equations of motion, $\rho=M \delta^{(d-1)}(r)$, therefore (we can relate the variations of the vierbein fields and those of the metric as $\delta e^{a} = \frac12 \delta g_{\mu\nu}\,g^{\nu\rho}\,e^a_\rho\,e^\mu_b\,e^{b}$)
\begin{equation}
\mathcal{E}_0\wedge e^0 = 4\pi T^0_0 \quad \Rightarrow \qquad \left[\frac{d}{d\log r}+(d-1)\right]\frac{\kappa}{r^{d-1}}\sim \rho ~,
\label{handy}
\end{equation}
and, as the right hand side of the equation does not depend on the Lovelock theory we are considering, the left hand side cannot either. Thus, the relation between $\kappa$ and the mass must be the same as in Einstein-Hilbert gravity (\ref{mass}). 

For arbitrary dimension, the spherically symmetric solutions where found in \cite{Boulware1985a,Wheeler1986,Wheeler1986a} and their extension to planar and hyperbolic symmetry was given in \cite{Cai1999,Aros2001}. Notice that this same expression holds for black holes arising in another class of theories called quasi-topological gravities \cite{Myers2010b} which, contrary to what happens in Lovelock theories (\ref{maximalK}), display in principle an unbounded highest curvature power $K$. We will comment more on this below.

\subsection{Branches}

Notice that (\ref{eqg}) leads to $K$ different roots for every value of the radius and, thus, to $K$ different branches associated to each of the cosmological constants (\ref{cc-algebraic}) (some of them may be imaginary, though), in such a way that $g_i(r\rightarrow\infty)=\Lambda_i$. For instance, in Gauss-Bonnet (GB) gravity there are two branches that read ($c_2\equiv \lambda L^2$)
\begin{equation}
g_{(\pm)} = -\frac{1}{2 L^2 \lambda} \left( 1 \pm \sqrt{1 - 4 \lambda \left( 1 - \frac{\kappa L^2}{r^{d-1}} \right)} \right) ~,
\label{GBbranches}
\end{equation}
each one associated with a different cosmological constant. The discriminant is, in this case, $\Delta=1-4\lambda$, and so we need $\lambda\leq1/4$ in order to have real solutions. $\lambda=1/4$ is the critical value of the GB coupling for which both vacua degenerate. Only one of the solutions, $g_{(-)}$, is {\it connected} to the standard Einstein-Hilbert gravity, in the sense that it reduces to it when $\lambda\rightarrow 0$,
\begin{equation}
g_{(-)} \approx -\frac{1}{2 L^2 \lambda} \left( 1 - \left[ 1 - 2 \lambda \left( 1 - \frac{\kappa L^2}{r^{d-1}} \right) \right] \right) = -\frac{1}{L^2} \left( 1 - \frac{\kappa L^2}{r^{d-1}} \right) ~,
\end{equation}
while $g_{(+)}$ blows up in that limit. It will be referred to as the EH-branch. It can be seen that this is the branch corresponding to the intersection of $\Upsilon[g]$ with the vertical axis, $g = 0$. The $K$ different branches of (\ref{eqg}) are continuous functions of the radial coordinate, as long as the roots of a polynomial equation depend continuously on its coefficients \cite{Marden1949}, and $r$ enters monotonically in the zeroth order coefficient $ \tilde c_0(r)\equiv c_0 -\kappa/r^{d-1}$. When $r\to\infty$, eq.(\ref{eqg}) is nothing but the expression leading to the $K$ cosmological constants.

\begin{figure}[h]
\centering
\includegraphics[width=0.54\textwidth]{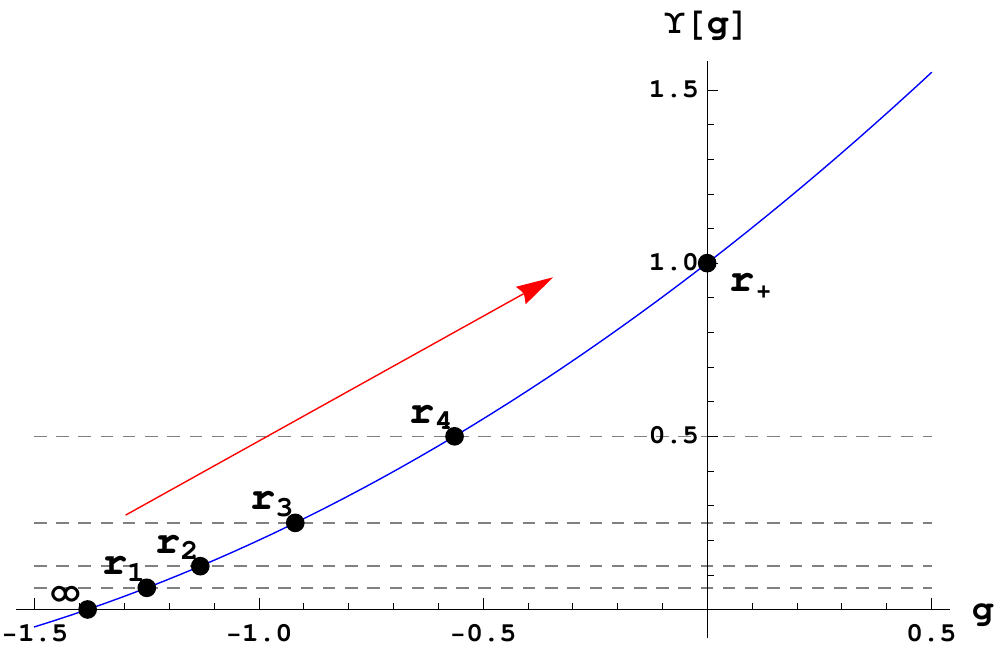}
\caption{A branch of the polynomial $\Upsilon[g]$ for the case $K=2$, {\it i.e.}, GB theory (with $\lambda = 0.2$ and $L = 1$), in arbitrary spacetime dimension, for different values of the radius ranging from $\infty$ to $r_+$, $r_1>r_2>\ldots>r_+$. The projection of the depicted points give $g(r_i)$ for the EH-branch in the planar case ($\sigma=0$).}
\label{polynomial-root}
\end{figure}

The different Lovelock couplings $c_{k>1}$ fix the shape of the polynomial $\Upsilon[g]$. While varying $r$ from $\infty$ to $r_+$ (see Figure \ref{polynomial-root}), the function $g(r)$ is given by the implicit solution of eq.(\ref{eqg}) that graphically corresponds to climbing up (down for negative masses) a given monotonic part of the curve $\Upsilon[g]$ starting from one of its roots (tantamount of a given comological constant). The same kind of analysis can be performed for charged solutions \cite{Camanho-u}, just by allowing the corresponding extra term in the right hand side of (\ref{eqg}).
The metric function $g$ is a monotonic function of $r$ since $\tilde c_0(r)$ is so and the remaining coefficients are frozen. Then each branch can be identified with a monotonic section of the polynomial $\Upsilon[g]$, and can easily be visualized graphically.

The propagator of the graviton corresponding to the vacuum $\Lambda_i$ is proportional to $\Upsilon'[\Lambda_i]$ in such a way that when $\Upsilon'[\Lambda_i]<0$ it has the opposite sign with respect to the Einstein-Hilbert case and thus the graviton becomes a ghost. This generalizes the observation first done by Boulware and Deser \cite{Boulware1985a}, and, because of it, we will be just considering positive slope branches. See \cite{Charmousis2008a} for a recent discussion on the subject. All the {\it relevant branches} correspond then to positive slope sections of the polynomial and, therefore, $g$ will be considered a monotonically decreasing function of $r$.

For positive $\kappa$ the solution runs over the points with positive value for $\Upsilon[g]$ while for negative mass it is the other way around. Either way, every branch always encounters a maximum/minimum, or it grows unboundedly.

For the sake of clarity and the ease of reading, let us first classify the different types of branches that one may encounter when dealing with a Lovelock theory of gravity. The appearance of a given type of branch will depend, in general, on the specific theory considered and on the values of the different coupling constants. On the one hand, we may classify the branches by their asymptotics: AdS, flat or dS branches. In the particular case we are considering, with $c_0=L^{-2}$, there are no asymptotically flat branches. The sign of the cosmological constant corresponding to the EH-branch (when real) is the opposite to $c_0$ (or, equivalently, the same as the explicit cosmological constant, as in standard Eintein-Hilbert gravity); thus, the EH-branch is asymptotically AdS. Due to the particular features and relevance of this branch, we will consider it separately.

Some of the branches (monotonic sections of the polynomial) may also be associated to complex values of $\Lambda$. Therefore, they do not correspond to real metrics and should be disregarded as unphysical. We will refer to these as excluded branches, and to the sector of the parameter space where the EH-branch is excluded as the {\it excluded region}. This region has been originally found for the cubic case in \cite{Camanho2010a} (see also \cite{Boer2009a}).

We will then exhaustively classify branches on (non-EH) AdS ({\it i.e.}, not crossing $g=0$), EH, dS and excluded branches. The latter, being unphysical, do not need further discussion. The AdS-branches must end at a maximum of the polynomial in order not to cross $g=0$. The other two cases may end at a maximum or, else, continue all the way up to $g\rightarrow\infty$. We will then consider two subclasses of branches: those (a) continuing all the way to infinity or (b) ending at a maximum. For the AdS-branches we will also consider two subclasses: (a) positive mass and (b) negative mass.

\subsection{Singularities and horizons}

Where are the singularities of these spacetimes located? The simplest way to answer this question is to calculate the curvature scalar and see where it diverges. As it depends on the metric and its derivatives, these divergences can be traced back to those of the first derivative of $g$,
\begin{equation}
g' = - \frac{(d-1)\kappa}{r^d} \Upsilon'[g]^{-1} ~.
\label{singularity}
\end{equation} 
Then, the metric is regular everywhere except at $r=0$ and at points where $\Upsilon'[g]=0$; that is, whenever the branch we are looking at coincides with any other. In such case, 
\begin{equation}
\Upsilon'[g]=\sum_{k=1}^{K} k\,c_k\, g^{k-1} = 0 ~.
\end{equation}
These are precisely the maxima/minima at which all branches end, except those growing unboundedly (that also approach asymptotically to a singularity located at $r = 0$). The values of $r$ where this happens exhibit a curvature singularity that prevents from entering a region where the metric becomes complex. This singularity rules out from scratch the possibility of solutions composed of different branches working at different intervals of the radial variable. 

It can be easily seen that the mass parameter $\kappa$ must be positive in the planar case ($\sigma=0$) in order for the spacetime to have a well defined horizon.  We can actually rewrite eq.(\ref{eqg}) as
\begin{equation}
\sum_{k=1}^{K} c_k\,  g^k = \frac{\kappa}{r^{d-1}} - \frac{1}{L^2} ~,
\end{equation} 
and realize that the equation admits a vanishing $g$ only when $r = r_+ \equiv (\kappa L^2)^\frac{1}{d-1}$. In the planar case, furthermore, only one branch has a horizon at $r_+$ and all the rest display naked singularities. This is so since the polynomial root $g=0$ has multiplicity one at $r=r_+$ (higher multiplicity would require a vanishing coefficient of the Einstein-Hilbert term). In the case of GB theory, for instance, we can see from eq.(\ref{GBbranches}) that it is $g_{(-)}$. This is the above mentioned EH-branch, a deformation of the solution to pure Einstein-Hilbert theory, and the only branch that remains when turning all the extra couplings off. Since $\Upsilon'[0] > 0$, $g(r)$ in the EH-branch is decreasing close to $r_+$ and, thus, it is a relevant branch.

For non-planar horizons the situation is more complicated and, in principle, some of the branches admit horizonful black hole solutions even for negative values of $\kappa$. The physical mass of the black hole has to match the one of the matter contained in that region of spacetime. Thus, $\kappa$ will be considered a positive quantity, except for hyperbolic black holes for which some comments on negative mass solutions shall be made. In \cite{Mann1997a}, indeed, the formation by collapse of black holes with negative mass has been considered. We shall see that spherical or planar black holes always exhibit a naked singularity in the case of negative mass. The only horizon that may arise for those solutions is a cosmological one. This is the case for negative mass asymptotically dS branches even for hyperbolic topology.

Taking into account that the value of $g$ at the event horizon, $r = r_+$, reads
\begin{equation}
g_+\equiv\frac{\sigma}{r_+^2} ~,
\label{gplus}
\end{equation}
we can write $\kappa = r_+^{d-1}\Upsilon[\sigma/r^2_+]$. The other way around, the radii of the location of the horizons are given by solutions of the previous equation for any given value of $\kappa$. This leads to a more handy formula for the mass
\begin{equation}
M = \frac{(d-2)V_{d-2}}{16\pi G_N}\, r_{+}^{d-1}\;\Upsilon\left[\frac{\sigma}{r_+^2}\right] ~.
\label{mass}
\end{equation}
Following the argument used to derive eq.(\ref{handy}), and taking into account that Einstein-Hilbert gravity has a positive energy theorem, we are tempted to conjecture that the same should apply to any Lovelock theory though this has not been proven so far. This conjectured positivity would not in principle rule out the negative mass solutions mentioned before as it may happen that the {\it positive mass} corresponds to the difference of the mass previously defined with the extremal one \cite{Horowitz1998}. 

We can recast the equation for the horizon, by means of (\ref{eqg}) and (\ref{gplus}), in such a way that it can be plotted in the $(g,\Upsilon[g])$ plane,
\begin{equation}
\Upsilon[g_+] = \kappa\left(\sqrt{\frac{g_+}{\sigma}}\right)^{d-1} ~,
\label{rheq}
\end{equation}
where the right hand side is just defined for positive/negative values of $g_+$, for $\sigma = \pm 1$, while for $\sigma=0$ the expression is not strictly valid since, in that case, $g_+$ exactly vanishes. Notice that in the high mass limit, $\kappa\rightarrow\infty$, the curve (\ref{rheq}) approaches the vertical axis --the planar black hole-- regardless of the value of $\sigma$.

\begin{figure}[h]
\centering
\includegraphics[width=0.57\textwidth]{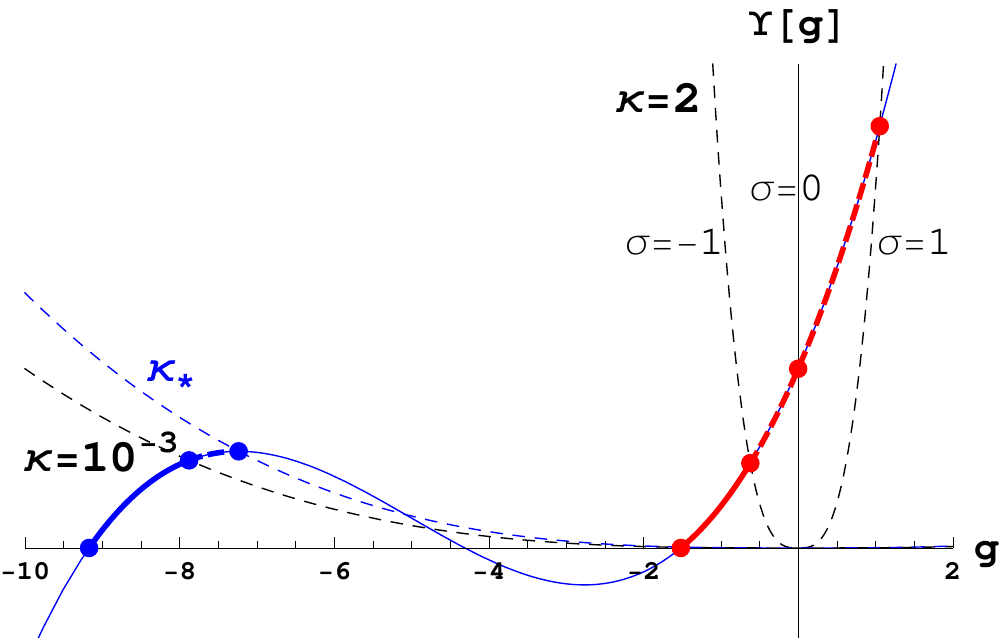}
\caption{Seven dimensional cubic Lovelock theory for $\lambda=1/4$ and $\mu=1/20$ ($L=1$) possesses two hyperbolic black holes for sufficiently low positive values of the mass. The dashed lines are plots of (\ref{rheq}) for the indicated values of $\sigma$ and $\kappa$ (in units of $L$). The crossing of these lines with the polynomial give the possible values for $g$ at the horizon and then of $r_+$. For $\sigma=-1$ and large enough values of $\kappa$, $\kappa > \kappa_\star$, the blue branch has a naked singularity (as it always has for $\sigma=0, 1$).}
\label{horizons}
\end{figure}

It is interesting to note that monotonicity of the function $g$ implies that every branch of black holes, for (positive mass) hyperbolic or planar topology, can have just one horizon. For $\sigma=0$ we recover just $g_+=0$, but for $\sigma\neq0$ we can actually have several possible values for $g_+$ (see figure \ref{horizons}).
Nonetheless, the right hand side of (\ref{rheq}) is monotonic in $g_+$ and each branch corresponds to a monotonic part of the polynomial $\Upsilon[g]$. We observe that, contrary to what happens for planar topology, there exists the possibility of having several branches with a horizon for $\sigma\neq0$. Some of them may be discarded by means of Boulware-Deser-like instabilities, while for some other branches horizons will appear or disappear depending on the actual values of the different couplings and $\kappa$.

For the case of hyperbolic horizons, as the slopes of both sides have opposite signs, there can just be at most one horizon per branch. In the positive curvature case, the determination of the number of horizons is, however, a non trivial matter. As the slope in both sides of the equation are positive we can even conceive the possibility of them crossing each other several times. We will illustrate this phenomenon below.

\begin{figure}[h]
\centering
\includegraphics[width=0.7\textwidth]{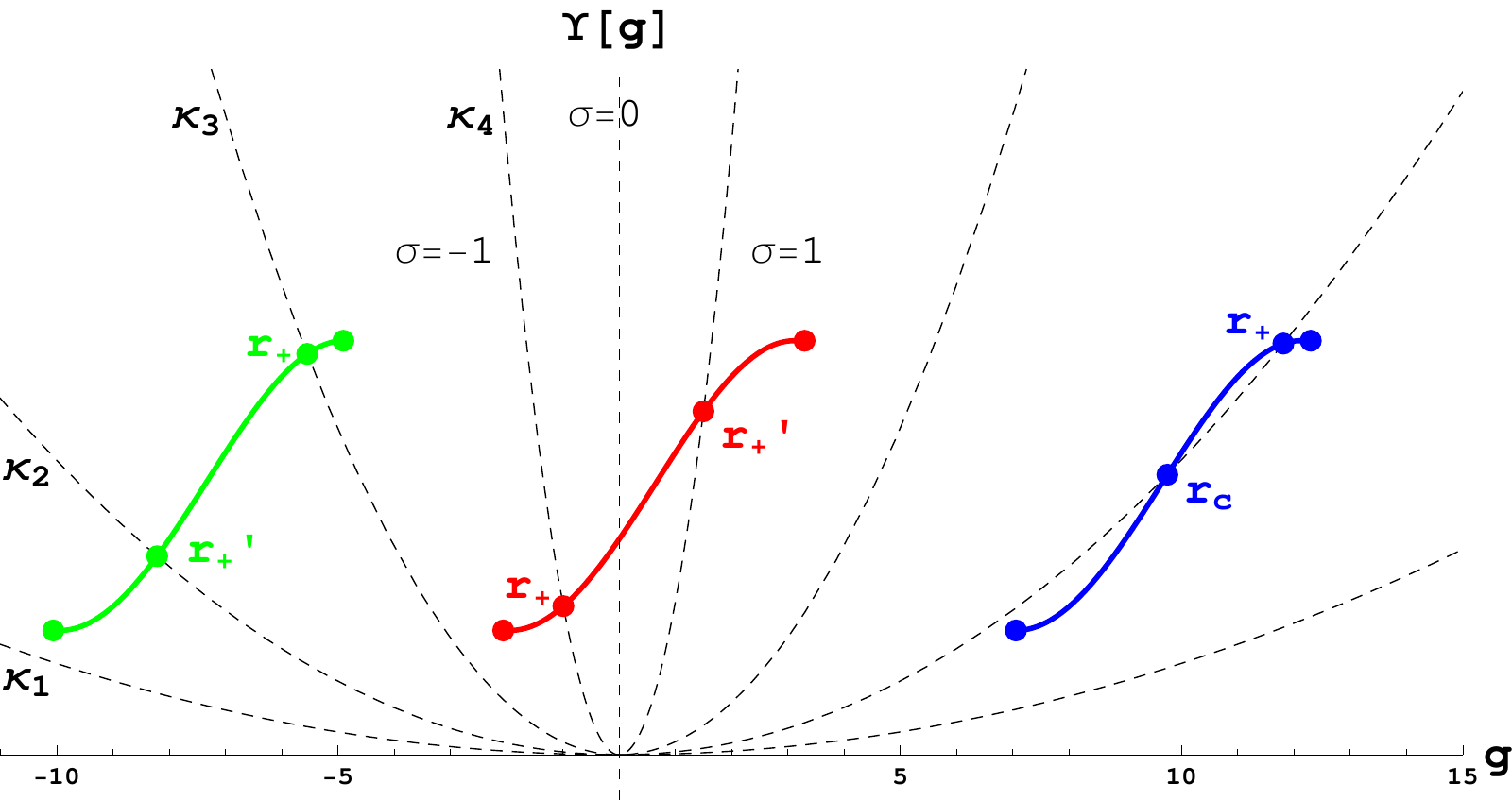}
\caption{Three examples of excluded branches running over positive, negative and positive, and just negative values of $g$ respectively. We also plot the values of the horizons for several values of the mass $\kappa_1<\kappa_2<\kappa_3<\kappa_4$ and all the topologies. The one crossing $g=0$ corresponds to the (excluded) EH-branch. The blue branch describes a well defined spacetime for some values of the mass with both singularities hidden behind the black hole and the cosmological horizons.}
\label{excluded}
\end{figure}

Depending on the couplings of Lovelock theory, it may happen that certain branches do not correspond to a proper vacuum. These coefficients fix the shape of the polynomial and, as they vary, some branches can become pathological in reason of their cosmological constant becoming imaginary. This happens whenever a monotonic part of the polynomial ends (towards the left) at a minimum without ever touching the $g$-axis (see figure \ref{excluded}). We refer to them as {\it excluded branches}. When the EH-branch is excluded we say that we are in the excluded region of the parameter space.
These spacetimes have two singularities, one for small values of the radial coordinate at the maximum, and another one for large values of $r$ at the minimum. In the cases where we can just have one horizon, the nakedness of the singularity associated with the minimum cannot be avoided. In the $\sigma=1$ case we may have two (or more) horizons, each of them hiding a singularity and describing a regular spacetime in between.

At this point it should be clear that several different kinds of branches may generically arise in Lovelock theory, depending on the topology of the spacetime slicing, the coupling constants and the relevant AdS/dS vacuum. These are schematically summarized in Table \ref{cases}.
\begin{table*}[h]
\centering
\begin{tabular}{c||c|c|c|}
asympt. & $\sigma=-1$ & $\sigma=0$ & $\sigma=1$ \\
\hline
\raisebox{9.3ex}{AdS}  & \includegraphics[width=0.26\textwidth]{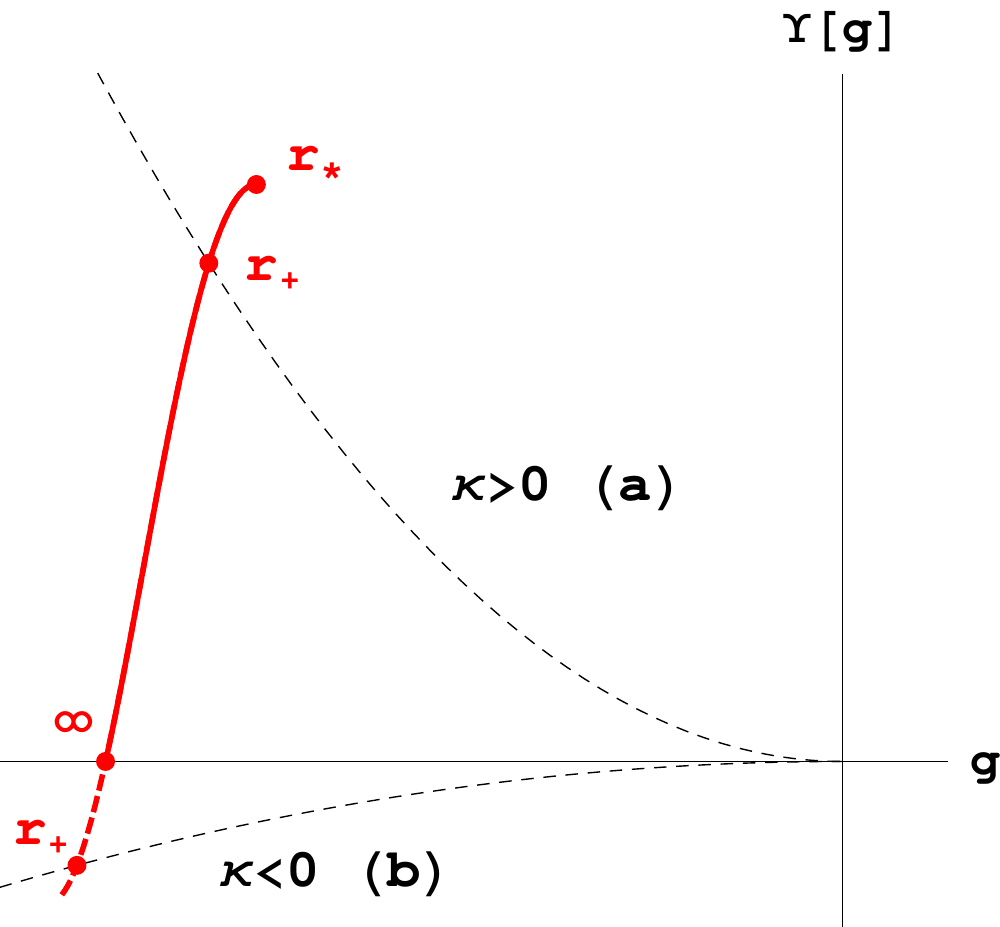} & \includegraphics[width=0.26\textwidth]{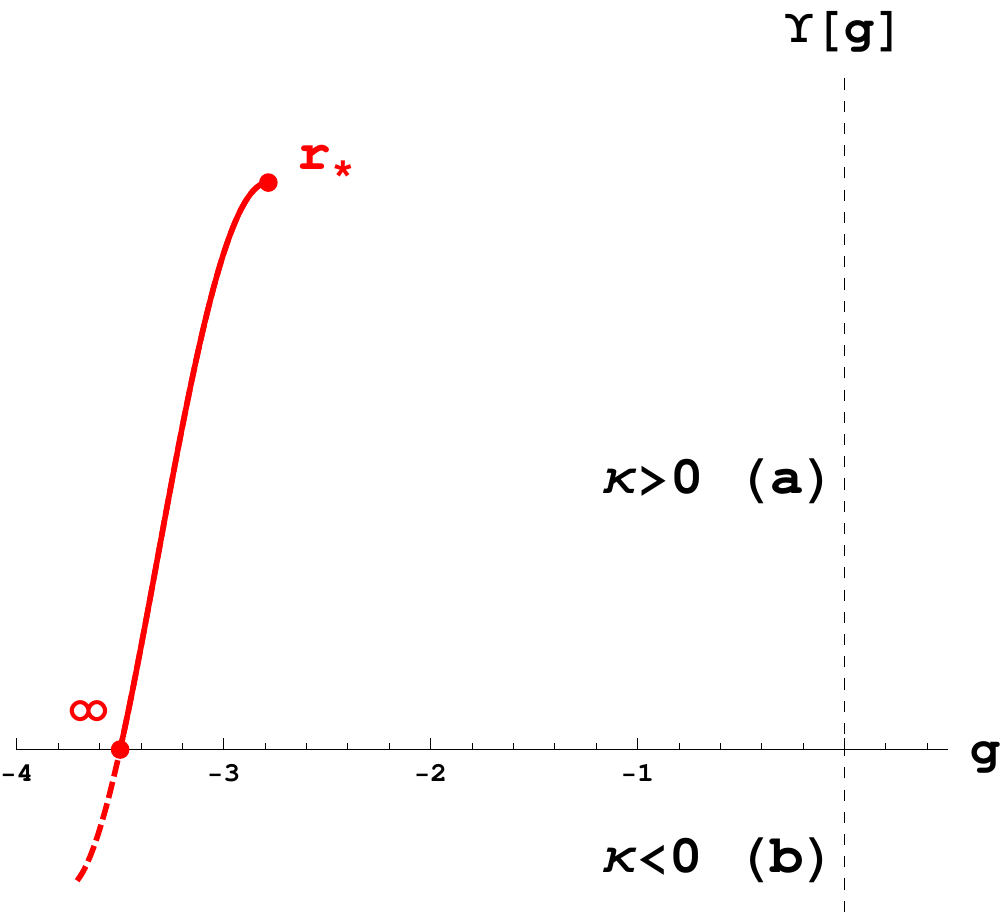} & \includegraphics[width=0.26\textwidth]{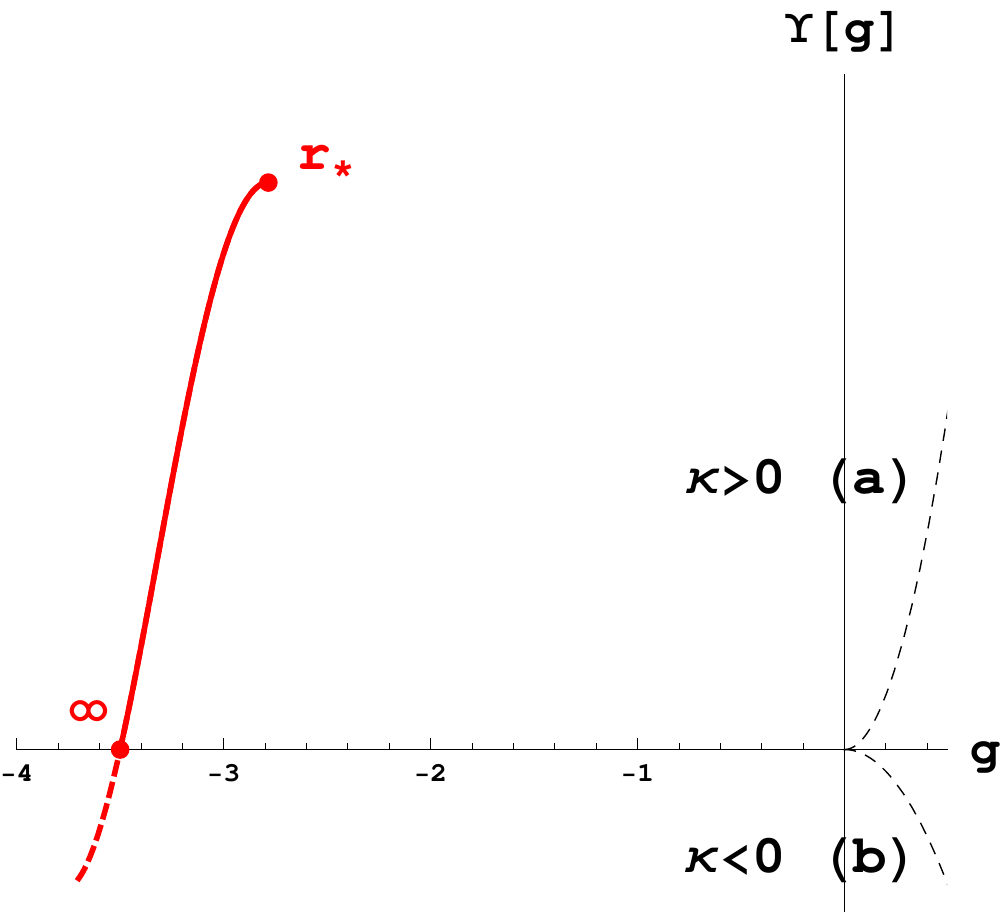} \\
\hline
\raisebox{9.3ex}{EH}  & \includegraphics[width=0.26\textwidth]{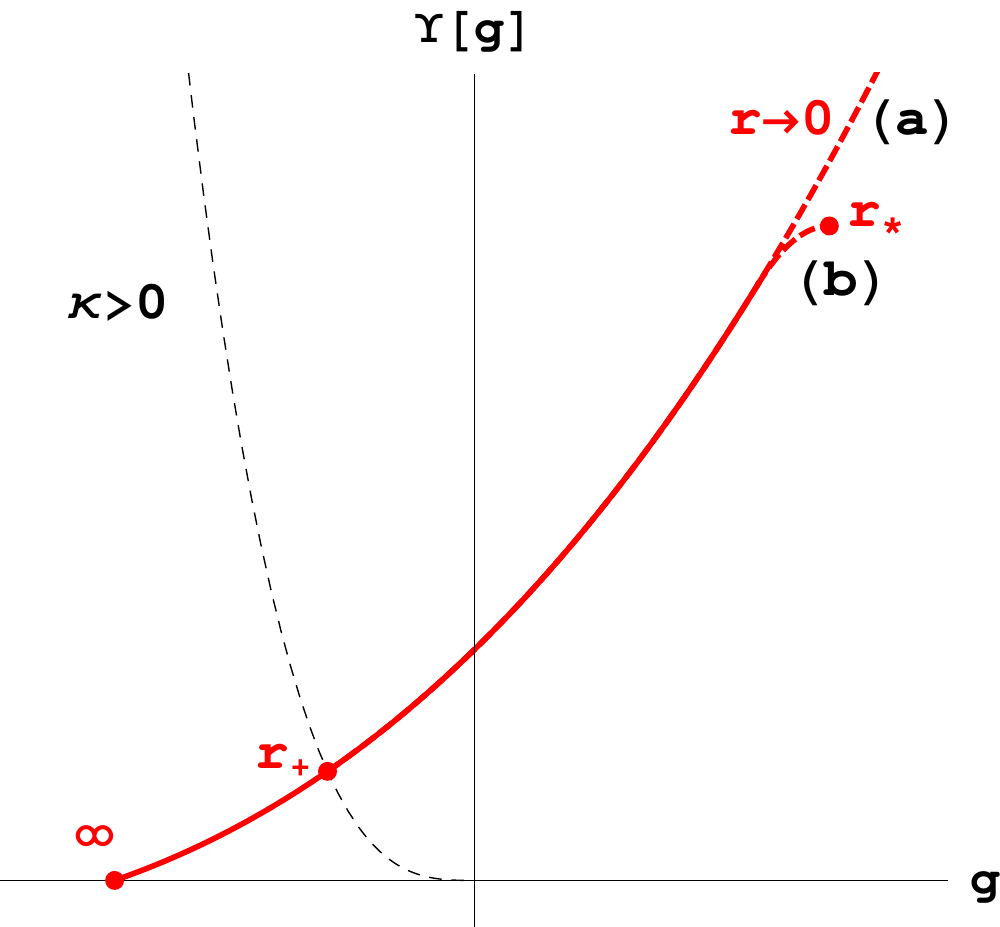} & \includegraphics[width=0.26\textwidth]{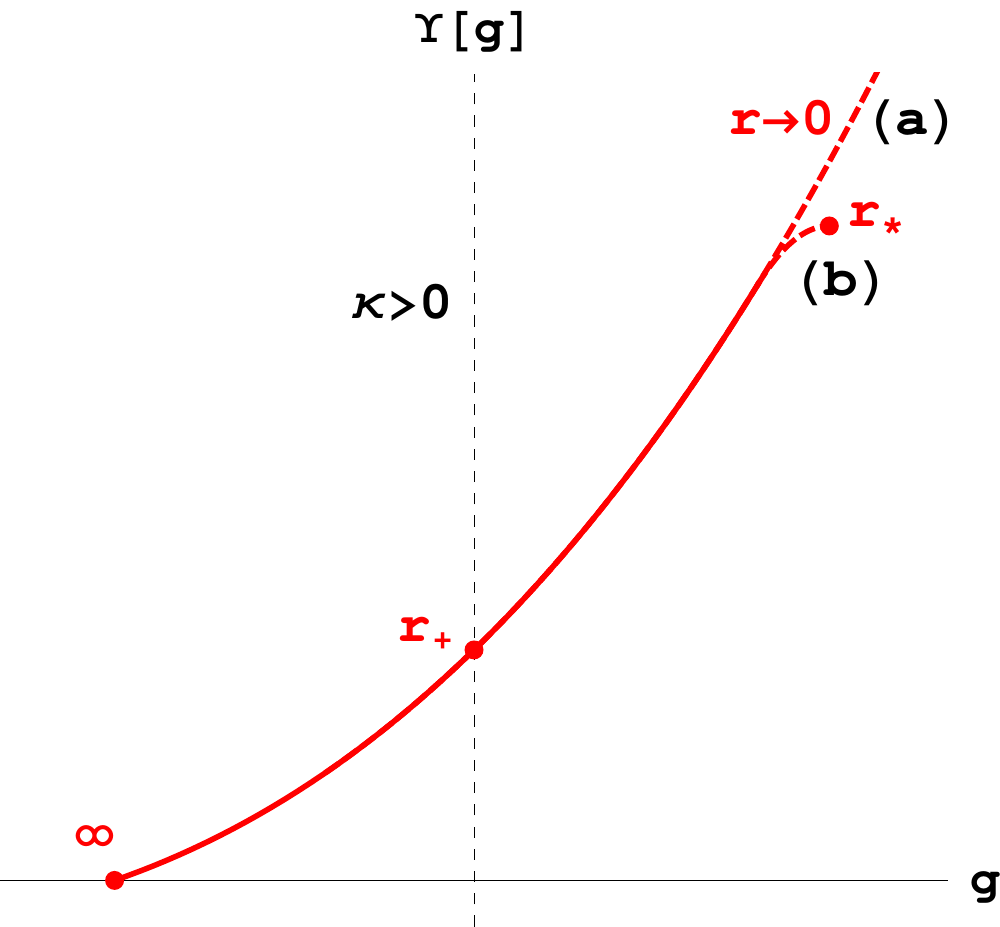} & \includegraphics[width=0.26\textwidth]{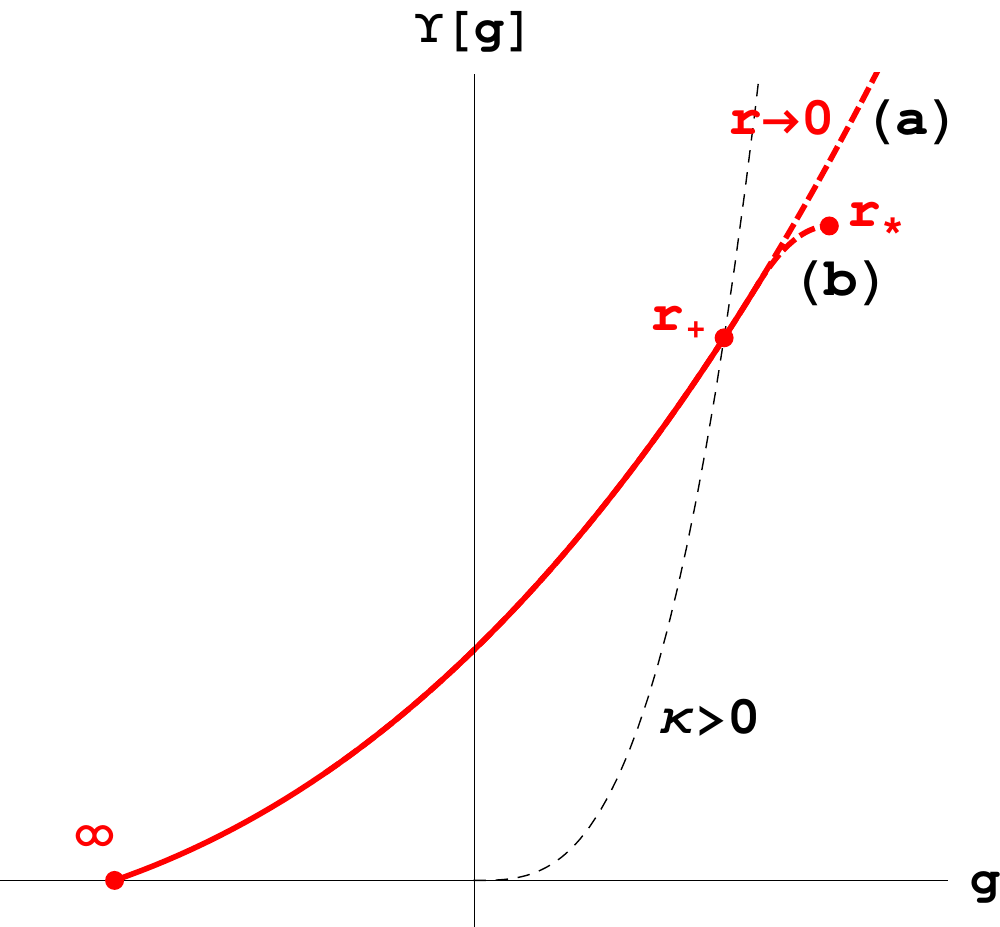} \\
\hline
\raisebox{9.3ex}{dS}  & \includegraphics[width=0.26\textwidth]{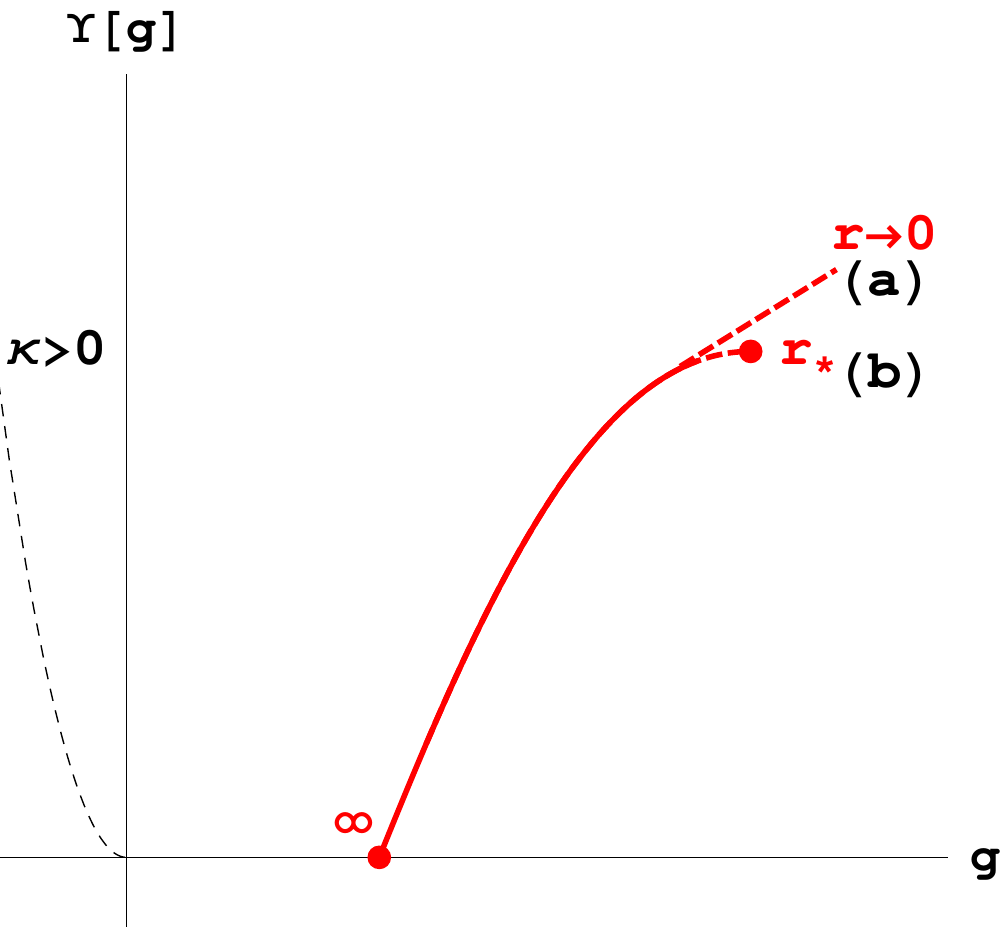} & \includegraphics[width=0.26\textwidth]{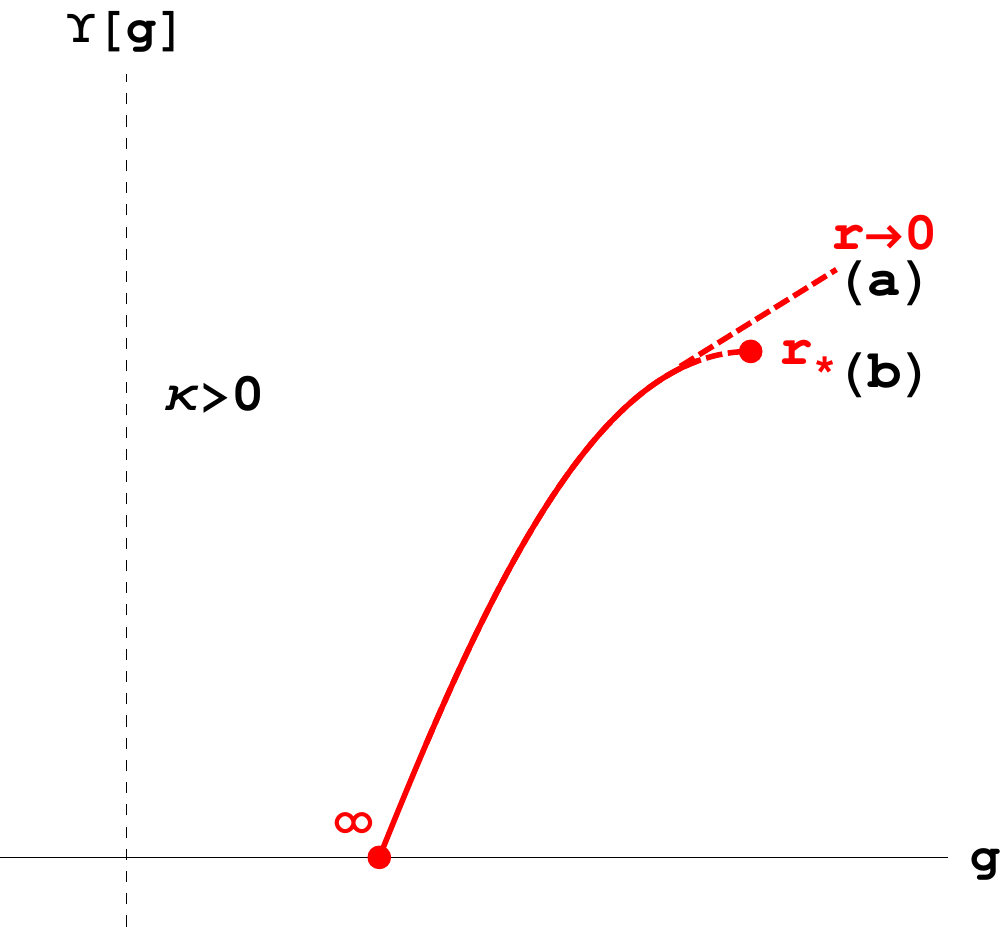} & \includegraphics[width=0.26\textwidth]{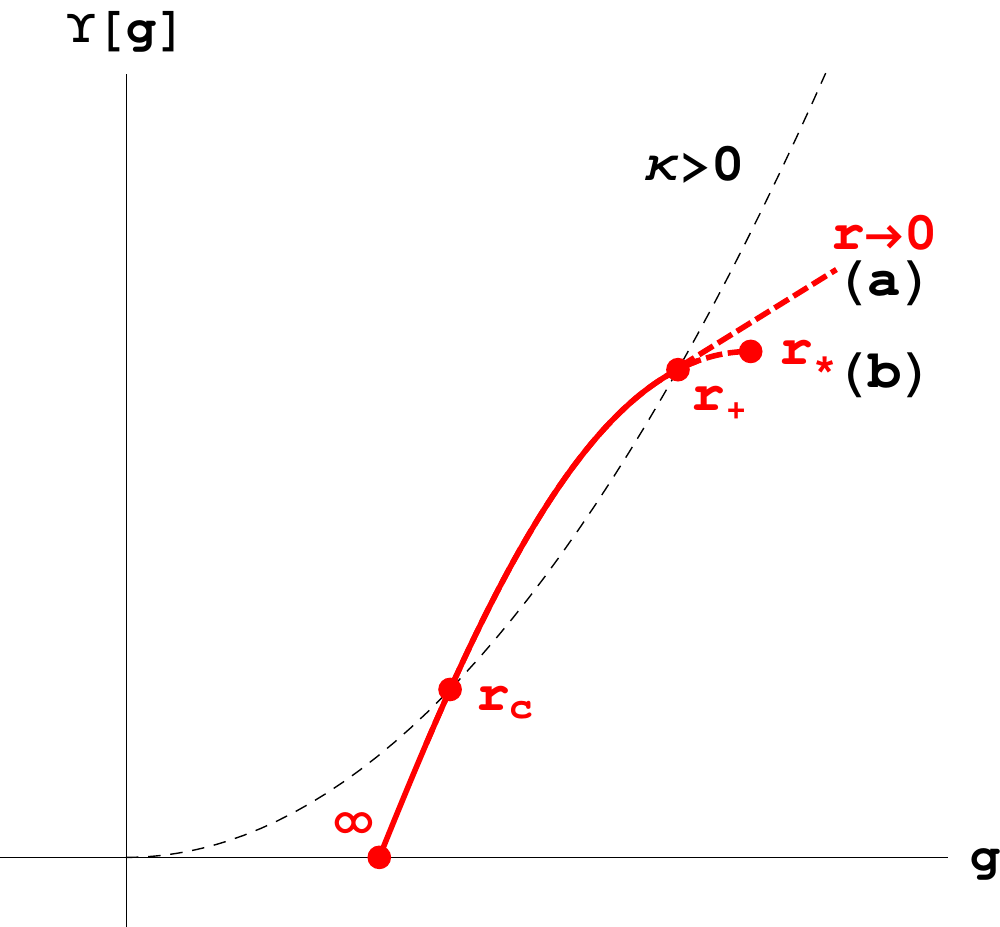} \\
\hline
\end{tabular}
\caption{Classification of (non-excluded) branches attending their asymptotics and topology. Black hole solutions exist in those cases where the given branch (in red) intersects the dashed curve: hyperbolic for an AdS-branch (top left), spherical for a dS-branch (bottom right), while only for the EH-branch supports all possible horizon topologies (second row). Spherical black holes in dS-branches exhibit, in addition to the event horizon, a cosmological horizon, $r_c$. We distinguish those branches ending up at extrema of $\Upsilon[g]$, called type (b), from those extending all the way to $r = 0$, named type (a).}
\label{cases}
\end{table*}

The existence of at least one horizon fixes hyperbolic topology as the only possible one for AdS branches (see the first row in the table), as well as it sets an upper bound on the mass of such spacetimes (corresponding to $r_\star$ in such plot). It also sets a lower bound if we consider the possibility of negative mass black holes in those branches. Also this sets a lower bound for the spherical black hole in the EH-branches that end up at a maximum, that we call type (b) (or even in those extending all the way to $r = 0$, named type (a), in the critical case, $d=2K+1$).

For dS branches this requirement also fixes the only possible topology admitting an event horizon as spherical at the same time as it imposes a double bound, upper and lower, as will be discussed further later on. The {\it physical} or untrapped region of the spacetime ($f>0$) is that located to the left of the dashed line in all figures appearing in the table. The region to the right corresponds to the inside of the would be horizon or trapped region ($f<0$).

\begin{figure}[h]
\centering
\includegraphics[width=0.43\textwidth]{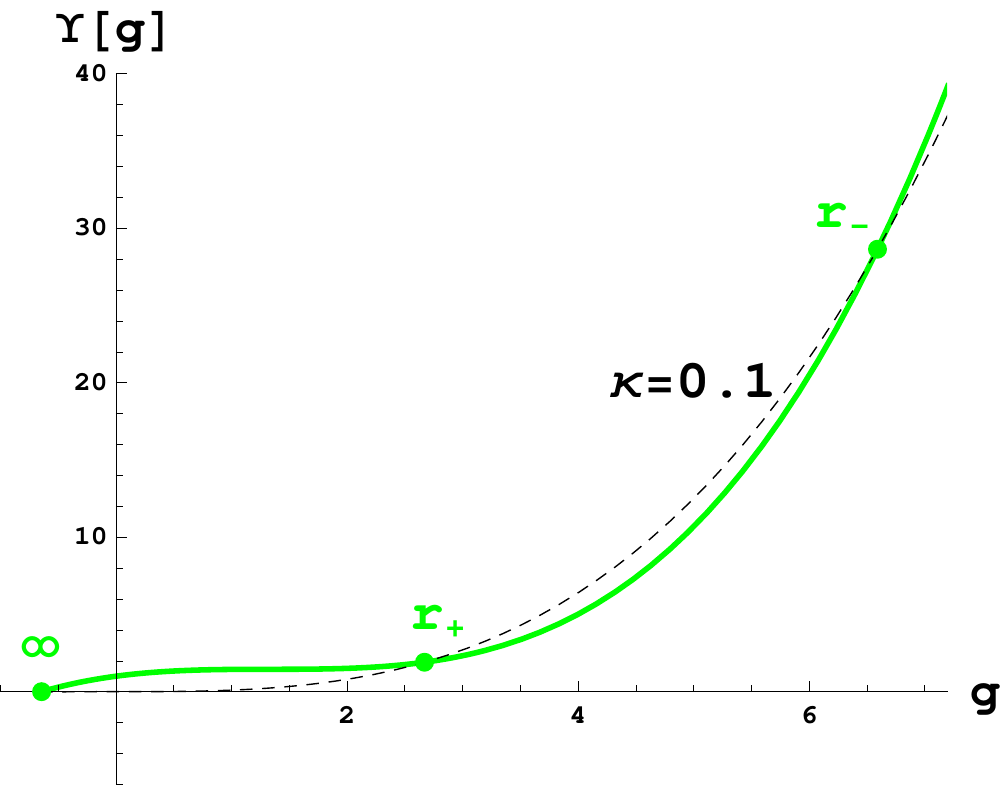}
\caption{Seven dimensional cubic Lovelock theory for $\lambda=-0.746$ and $\mu=0.56$. In the green branch we observe the occurrence of two (outer and inner, respectively) horizons, $r_+$ and $r_-$. For $d=8$ we find a similar behavior with one further inner horizon, three in total.}
\label{mhorizons}
\end{figure}

We will also have more untrapped regions inside the black hole in the presence of several black hole horizons. We already mentioned, indeed, the possibility of black hole spacetimes with multiple horizons. This is for instance the case for some regions of the parameter space with (either EH or dS) branches displaying inflection points. The simplest situation where this can be observed is therefore the cubic theory, as shown in figure \ref{mhorizons}.
This very same behavior will be found in general for some region of the parameter space in the critical Lovelock theory; for $d=2K+1$ this can be easily understood as we can always construct a polynomial 
\begin{equation}
\Upsilon[g]=\prod_{i=1}^{L\leq K} {\left(1-\frac{g}{g_i}\right)} + \alpha\,g^K ~,
\end{equation}
for large enough $\alpha$ and suitable coefficients, $g_i > \Lambda_\star$, in order to make the slope everywhere positive for the EH or dS branches with cosmological constant $\Lambda_\star$. Then, for $\kappa=\alpha$, the polynomial equation has all $g_i$ as solutions. Different (positive) $g_i$ correspond to different spherical horizons, each degenerate $g_i$ giving rise to a degenerate horizon. From this value, varying the mass of the solution the number of horizons will in general change. For the EH-branch, the number of black hole horizons has to be always one for high enough mass, and it is so as well in the low mass regime for $d>2K+1$ as well for the EH as for any dS branch. Thus, we have in general couples of horizons appearing and disappearing depending on the values of the mass. We will always refer as $r_+$ to the outermost horizon of the black hole, {\it i.e.}, the biggest root of (\ref{rheq}) besides the cosmological horizon, if present. The very same logic applies to the case of negative mass black holes with hyperbolic horizon. We will comment more on this later.

\subsection{Thermodynamics}

Some aspects of Lovelock black holes thermodynamics have been considered earlier in \cite{Cai2004}. In the present subsection we will proceed to a complete analysis including all possible cases and branches.

The EH-branch is the only one that does not display naked singularities (at least for $\sigma=0$ or $-1$). For $\sigma=1$, naked singularities can be prevented by considering a big enough black hole. In that way we also ensure the stability of the black hole, as it happens for the usual Einstein-Hilbert AdS-Schwarzchild black hole. For large $r_+$, we can approximate
\begin{equation}
M \sim T^{d-1} ~,
\end{equation}
which coincides with the planar case. Then, $dM/dT > 0$ and the black hole is locally thermodynamically stable; it can be put in equilibrium with a thermal bath. In general, this will not happen for small black holes. This points towards the occurrence of Hawking-Page-like phase transitions of the kind interpreted as confinement-deconfinement phase transitions in the corresponding dual conformal plasma \cite{Witten1998,Witten1998a}, which have been already studied in the case of GB gravity \cite{Nojiri2001j,Cho2002a}. In order to further investigate these in the case of general Lovelock theories it would be necessary to calculate the free energy (the euclideanized on-shell action) associated with the black hole. This has to be studied in the non-planar case, otherwise scale invariance would only allow for phase transitions at zero temperature. The complicated structure of phase transitions  would then be hidden and any black hole will correspond to the stable highest temperature deconfined phase.

If we want to consider CFTs on flat spacetime, we just need to consider flat horizon black holes: {\it i.e.}, just the EH-branch already discussed in \cite{Camanho2010a}. If we plug the asymptotic behavior of $g$ at infinity, $g \approx \Lambda \left( 1 - \tilde{\kappa}/r^{d-1} \right)$, in (\ref{eqg}), we get
\begin{equation}
\tilde{\kappa} = - \frac{\kappa}{\Lambda\,\Upsilon'[\Lambda]} ~.
\label{kappatilde}
\end{equation} 
Thereby, as long as the polynomial factor is positive and the cosmological constant negative, $\tilde{\kappa}$ is positive: The well-behaved black hole associated with the EH-branch always behaves asymptotically as the usual (positive mass) AdS black hole solution of Einstein-Hilbert gravity. The addition of higher curvature corrections does not change the qualitative behavior of the solution in any meaningful way. These black holes have a well defined temperature
\begin{equation}
T = \frac{f'(r_+)}{4\pi} = \frac{r_+}{4\pi}\,\left[(d-1)\,\frac{\Upsilon[g_+]}{\Upsilon'[g_+]}-2\,g_+ \right] ~.
\end{equation}
From this expression, however, it is not clear what the sign of the temperature is. The first term is always positive, for positive slope branches and mass. The sign of the second term depends on $\sigma$ in such a way that it is negative for spherical topology. Then the temperature is trivially positive for positive mass black holes with hyperbolic or planar topology. It is also positive for negative mass AdS black holes as will be explained shortly. This simply derives from the fact that the temperature can just change sign if it vanishes at some intermediate $r_+$, {\it i.e.}, if the black hole is extremal and the event horizon degenerates ($f'=0$). This very same logic will ensure the positivity of the temperature of spherical black hole horizons. This will be explained later following a different approach.  

We will also have negative temperature horizons but this is a common feature of Einstein-Hilbert gravity. For instance, the temperature of the cosmological horizon of pure dS space is negative in our conventions, as it is negative for the inner horizon of a Reissner-Nordstrom black hole, regardless of the asymptotics of the solution. The same will happen here.

One important feature that will be relevant later on, when discussing classical stability, is that the temperature is proportional to the derivative of the mass (\ref{mass}) with respect to the radius of the black hole horizon,
\begin{equation}
\frac{dM}{dr_+} = \frac{(d-2)V_{d-2}}{4\, G_N }r_+^{d-3}\;\Upsilon'\left[g_+\right]\, T ~.
\label{dMdr}
\end{equation}
The proportionality factor is exactly the radial derivative of the black hole entropy
\begin{equation}
\frac{dS}{dr_+}=\frac{1}{T}\frac{dM}{dr_+}=\frac{(d-2)V_{d-2}}{4\, G_N }\,r_+^{d-3}\;\Upsilon'\left[g_+\right] ~.
\label{entropy}
\end{equation}
We can see that, as long as  we are in a branch free from Boulware-Deser instabilities, the radial derivative of the mass and the entropy are positive for $T>0$. On the one hand, this will be important when discussing classical instability as the heat capacity of the black hole reads
\begin{equation}
C=\frac{dM}{dT}=\frac{dM}{dr_+}\frac{dr_+}{dT} ~,
\end{equation}
and then the only factor that can be negative leading to an instability is $dT/dr_+$. On the other hand, if we set the ground state ($r_+=0$) entropy to vanish, the positivity of (\ref{entropy}) implies the positivity of the entropy. Negative values for the entropy may however be encountered \cite{Myers1988, Nojiri2001n, Cvetic2002, Neupane2009a, Neupane2009f} in the case of hyperbolic horizons in which such {\it ground state} does not exist (also for type (b) spherical solutions). It is actually unclear what is the suitable vacuum solution to be used as a {\it reference state} in those cases \cite{Emparan1999}.

The expression for the entropy can be easily obtained by integrating (\ref{entropy}). It is the only magnitude seen so far that cannot be readily expressed in terms of $\Upsilon[g]$,
\begin{equation}
S \equiv \frac{V_{d-2}}{4\, G_N }\,r_+^{d-2}\sum^{K}_{k=1}{k\,c_k\frac{d-2}{d-2k}\, g_+^{k-1}} =\frac{A}{4\,G_N}\left(1+\sum_{k=2}^{K}{k\,c_k\frac{d-2}{d-2k}\, g_+^{k-1}}\right) ~,
\end{equation}
and it coincides with the entropy obtained by other means like the Wald entropy \cite{Jacobson1993} or the euclideanized on-shell action \cite{Myers1988}. The resulting expression for $K=3$ coincides with that arising in cubic quasi-topological gravity \cite{Myers2010b}. We fixed the integration constant such that the entropy vanishes when the horizon radius goes to zero. For planar horizons ($g_+=0$) this formula reproduces the proportionality of entropy and area of the black hole horizon, $A=V_{d-2}\;r_+^{d-2}$, whereas it gets corrections for other topologies. Interestingly enough, it has been recently suggested that this expression can be extended towards the interior of the geometry by performing a radial foliation and replacing $g_+$ by $g(r)$; the resulting function, $S(r)$ being interpreted as the information contained inside a given region of the spacetime \cite{Paulos2011}. 

From these quantities we can now compute any other thermodynamic potential such as the Helmholtz free energy, $F=M-TS$, 
\begin{equation}
F=\frac{(d-2)V_{d-2}}{16\pi G_N}\frac{r_+^{d-1}}{\Upsilon'[g_+]}\sum_{k,m=0}^{K}{\frac{2m-2k+1}{d-2k}\,k\,c_k\,c_m\,g_+^{k+m-1}} ~.
\label{free-energy}
\end{equation}
This magnitude is relevant for processes at constant temperature to analyze the global stability of the solutions. The above expression has degree $2K-1$ in the numerator as a function of $g_+$. Thus, that is the maximal number of zeros that may eventually correspond to Hawking-Page-like phase transitions, provided they have the adequate sign for each topology (see \cite{Neupane2004} for a concrete example in the Gauss-Bonnet theory). This expression also diverges at a maximum of the polynomial. We will comment further about this below. 

\begin{figure}[h]
\centering
\includegraphics[width=0.44\textwidth]{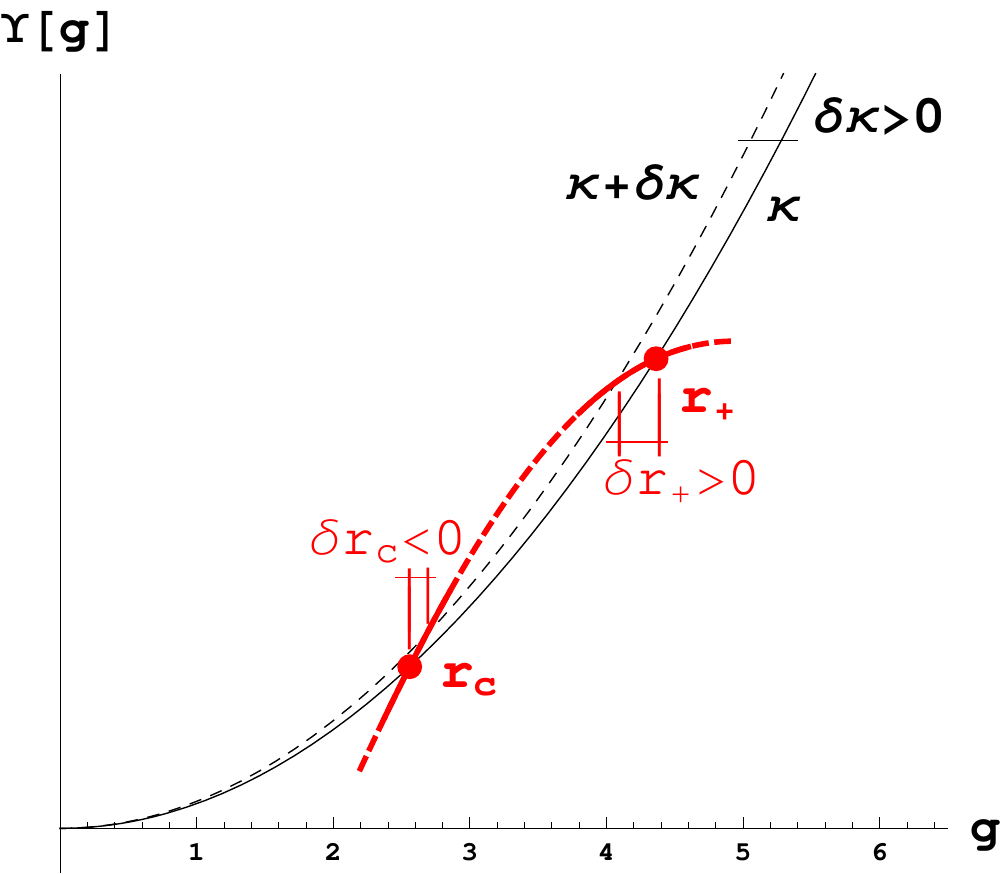}
\caption{Determination of the sign of the temperature for the cosmological and black hole horizon of a dS branch. The cosmological horizon has $T_c\propto d\kappa/dr_c <0$ whereas the event horizon of the black hole has $T_+\propto d\kappa/dr_+ >0$. Recall that ${\rm sign}(\delta g_+)=-{\rm sign}(\delta r_+)$ and the same holds for every horizon.}
\label{signT}
\end{figure}

Another useful application of (\ref{dMdr}) is to determine the sign of the temperature, which is the same as that of the variation of the mass with respect to the horizon radius. One can easily realize (see figure \ref{signT}) that the sign of the temperature depends just on the direction of the change of sign of the function $f$ across the horizon.
If this sign changes from $f<0$ to $f>0$ as, for instance, in the cosmological horizon of any asymptotically dS spacetime, the temperature is negative, whereas it is positive in the opposite case. The black hole horizon, as the largest root of (\ref{rheq}), always corresponds to the latter case, separating an untrapped region ($f>0$) from a trapped one ($f<0$), and as such it always has positive temperature. The inner horizons have alternating signs for the temperature starting from a negative one. Degenerate horizons obviously have zero temperature corresponding to extremal black holes. 

\subsection{Vacuum horizons and Einstein-Hilbert gravity}

Let us start this subsection by discussing the horizon structure of the vacuum solutions. The general form of the metric function $f$ is in this case
\begin{equation}
f(r)=\sigma-\Lambda\,r^2 ~,
\end{equation}
so it can vanish at $r=\sqrt{\sigma/\Lambda}$, whenever $\sigma$ and $\Lambda$ have the same sign, thus for hyperbolic AdS and spherical dS spacetimes. These horizons are observer dependent features since these spacetimes are maximally symmetric and, thus, any point can be considered as the origin. The dS case is widely known \cite{Gibbons1977a}, this corresponding to the cosmological event horizon.

The AdS case is, however, more obscure as long as the horizon is actually cloaking a finite size region in a similar way as a regular black hole horizon does. The black hole horizon is actually just a deformation of this `vacuum' horizon. This has a problematic interpretation and has led to the proposal that the true ground state for hyperbolic spacetimes is not the massless one, but an extremal negative mass solution \cite{Vanzo1997a, Birmingham1999}. The cosmological horizon of pure dS spacetime has negative temperature, while for the AdS case the temperature is positive as for a regular black hole horizon.  

\begin{figure}[h]
\centering
\includegraphics[width=0.58\textwidth]{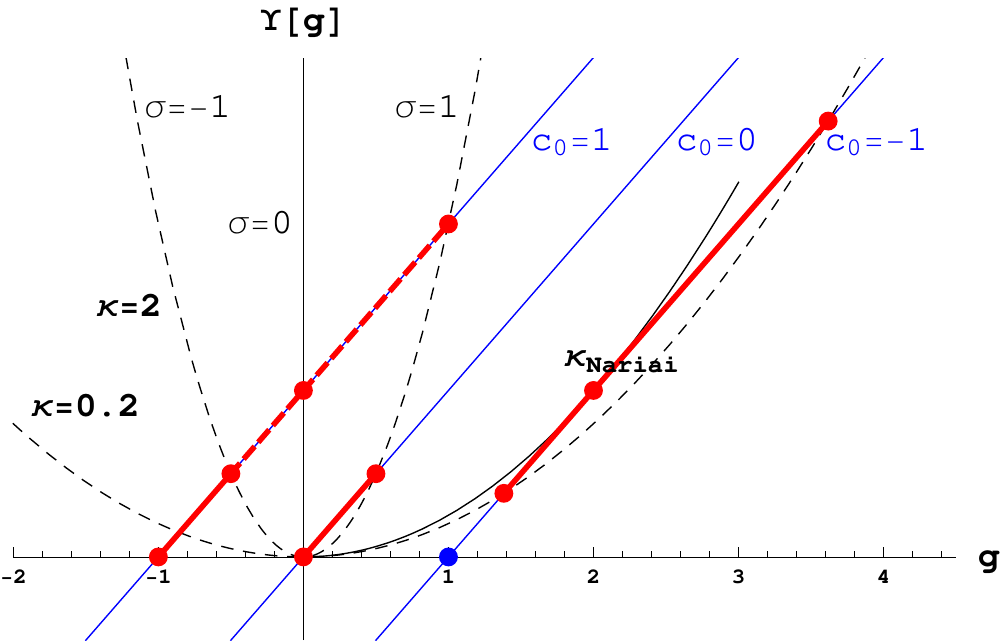} 
\caption{Linear polynomial corresponding to the usual EH-branch for negative ($c_0=1$), zero ($c_0=0$) and positive ($c_0=-1$) cosmological constants ($L=1$). The dashed lines are just $\kappa\left(g/\sigma\right)^{\frac{d-1}{2}}$ for $d=5$. The solid black line corresponds to the critical value of the mass, $\kappa_{\rm Nariai}=1/4$, for dS black holes with spherical horizon. The crossing of these lines with the polynomial gives the possible values for $g$ at the horizon and then of $r_+$ (and $r_c$). For $\sigma=1$ and $\kappa>\kappa_{\rm Nariai}$ (or $r_+>r_{\rm Nariai}$), the asymptotically dS branch describes a big crunch spacetime ($f<0$, $\forall r$) without horizons.}
\label{horizonsEH}
\end{figure}

In order to analyze the horizon structure, let us focus on the asymptotically AdS, dS and flat black holes in Einstein-Hilbert gravity with cosmological constant \cite{Birmingham1999}. We include this simple case here for completeness and as an illustration of our method. As clearly depicted in figure \ref{horizonsEH}, the only case accepting all three distinct topologies without exhibiting naked singularities is the asymptotically AdS configuration, the other two cases being well-defined just for spherical topology.
This AdS case, furthermore, has just one horizon for all three topologies. The asymptotically flat spherical black hole has one event horizon as well. The asymptotically dS spherical black hole has in general two horizons: One of them is just the deformation of the cosmological horizon already present in the maximally symmetric solution, while the other corresponds to the black hole. As the mass increases both horizons get closer to each other until, for some critical value of the mass, the so-called {\it Nariai mass},
\begin{equation}
\kappa_{\rm Nariai} = \frac{2L^{d-3}}{d-1}\left(\frac{d-3}{d-1}\right)^{\frac{d-3}{2}} ~,
\end{equation} 
they actually meet (they disappear for masses above that value). The untrapped region, the spacetime as we usually consider it, is comprised between the two horizons and so for this extremal case it seems to disappear. A proper limiting procedure \cite{Ginsparg1983} shows that the geometry remains perfectly regular as $\kappa \rightarrow \kappa_{\rm Nariai}$, and becomes the geometry of the Nariai solution. This space is the direct product of a dS$_2$ and a S$^{d-2}$, both with the same radius. Above this critical mass, though, it describes a {\it big crunch} spacetime.

\begin{figure}[h]
\centering
\includegraphics[width=0.43\textwidth]{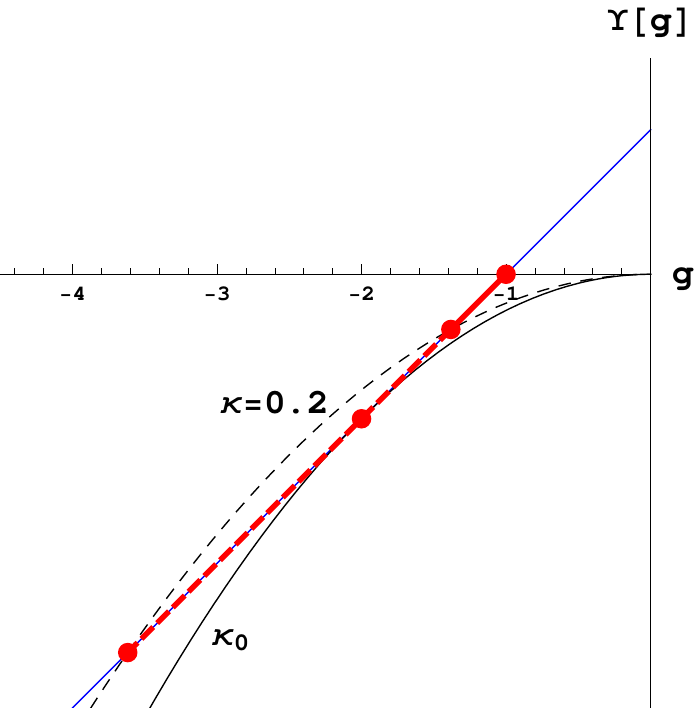} 
\caption{Negative mass hyperbolic black holes in Einstein-Hilbert gravity. The dashed line corresponds to a black hole with an outer and inner horizon (segment in red), while the solid line represents the extremal case, $\kappa_0=-\kappa_{\rm Nariai}$.}
\label{negM}
\end{figure}

In the asymptotically AdS case, a negative mass extremal hyperbolic black hole has been proposed as the ground state in Einstein-Hilbert gravity. The same would apply to any Lovelock theory. Black holes with larger (but negative) mass than this extremal one,
\begin{equation}
\kappa_0=-\kappa_{\rm Nariai}=-\frac{2L^{d-3}}{d-1}\left(\frac{d-3}{d-1}\right)^{\frac{d-3}{d-1}} ~,
\end{equation}
have two horizons, in a way reminiscent of the asymptotically dS spherical black hole with positive mass. The difference being that in the asymptotically dS case the two correspond respectively to the cosmological and black hole horizons, while in the AdS case they are the outer and inner horizons of a black hole (see figure \ref{negM}).
It is worth noticing that for negative masses we are exploring a completely different section of the polynomial than for positive masses, the similarities being just due to the extremely simple form of the polynomial in the case under current analysis. In general, positive and negative mass black holes may have dramatically different features.

As we can easily see in this simple example, the existence of just one black hole horizon in all positive mass cases implies that the singularity at the origin is always spacelike, while it is timelike in the negative mass hyperbolic case due to the presence of two black hole horizons that become degenerate in the extremal limit. In the case where the horizon and the singularity coincide, the nature of the latter is light-like.

\section{Taxonomy of Lovelock black holes}

We will study generic features of maximally symmetric Lovelock black holes in a case by case basis, considering the previously introduced classes of black hole branches (table \ref{cases}). Some work in this direction has already been done considering just the Gauss-Bonnet case \cite{Torii2005,Torii2005a}. Let us have a look on the different cases that we can encounter.

\subsection{The Einstein-Hilbert branch}

The Einstein-Hilbert branch is just a deformation of the Schwarzschild-AdS black hole ($c_0=L^{-2}$) and can be identified as the branch crossing $g=0$ with slope $\Upsilon'[0]=1$, exactly as in the Einstein-Hilbert case, and so the slope will be positive for the whole branch. This condition protects this branch from Boulware-Deser-like instabilities. When real, the cosmological constant associated with this branch is negative and so the spacetime is asymptotically AdS. We can proceed with this analysis in an analogous way for asymptotically dS spaces, just by changing the sign of the explicit cosmological constant in the action $c_0\rightarrow -\frac{1}{L^2}$, or for asymptotically flat ones, just by setting $c_0=0$. We include the relevant part of table \ref{cases} below, for the reader convenience.

\begin{table*}[h]
\centering
\begin{tabular}{c||c|c|c|}
asympt. & $\sigma=-1$ & $\sigma=0$ & $\sigma=1$ \\
\hline
\raisebox{9.3ex}{EH}  & \includegraphics[width=0.26\textwidth]{EHh} & \includegraphics[width=0.26\textwidth]{EHf} & \includegraphics[width=0.26\textwidth]{EHs} \\
\hline
\end{tabular}
\caption{Taxonomy of the EH-branch black holes.}
\label{EH-case}
\end{table*}

Even though the EH-branch is just a deformation of the usual Schwarzschild-AdS black hole, it can be a quite dramatic one. For instance, it may happen that the polynomial has a minimum at $g_{\rm min} < 0$ (if there are several, $g_{\rm min}$ refers to the lowest one in absolute value), such that $\Upsilon[g_{\rm min}] > 0$. A naked singularity would arise for large radius: the solution does not approach AdS asymptotically. This case was first discussed in \cite{Boer2009a,Camanho2010a} for third order Lovelock theory and planar topology, but the same applies in the general case for a vast region of the space of parameters that will be named, following the aforementioned reference, the {\it excluded region}. In order to avoid the excluded region the value of $\Upsilon[g_{\rm min}]$ at the biggest negative minimum, $g_{\rm min}$, has to be negative. The sector of the parameter space where this new kind of singularity appears has to be excluded in general, not only because of its nakedness but in reason of the perturbative instability of the corresponding solution \cite{Camanho2012}. 

For {\it hyperbolic} or {\it planar} topology, as this branch always crosses $g=0$ with positive slope, $\Upsilon'[0]=1$, it has always a horizon hiding the singularity of the geometry that is located (see table \ref{EH-case})
\begin{itemize}
\item
either at $r=0$ [(a) type],\vskip2mm
\item
or at the value corresponding to a maximum of $\Upsilon[g]$ [(b) type].
\end{itemize}
For hyperbolic horizons we have again the possibility of considering negative mass black holes, for masses above a critical value corresponding to the extremal case. This makes no difference with respect to the same situation taking place in an AdS branch and, thus, will be discussed at length below.

The {\it spherical} case is quite more involved. For high enough mass, the existence of the horizon is ensured, but this is not the case in general. For the (a) type EH-branch the existence of the horizon can be elucidated by analyzing (\ref{rheq}) in the limit of small mass $g_+\rightarrow\infty$. For the horizon to exist in this limit, we need the right hand side of the equation to be bigger than the left hand side. This is ensured for $d>2K+1$ as in this case the biggest power in the left hand side would be smaller than $(d-1)/2$. The existence of a horizon in the small mass limit ensures the existence of (at least) one for all values of the mass, simply due to the continuity of $\Upsilon[g]$.

The case $d=2K+1$ is {\it critical}. There will be a minimal mass ($\kappa_{\rm crit} \leq c_K$) below which a naked singularity appears. In principle, for high enough orders of the Lovelock polynomial, more than one horizon can exist but for the critical case, at some point, all of them disappear. The number of black hole horizons determines the type of singularity situated at $r=0$, space or timelike. For $d>2K+1$ we will always have an odd number of horizons (taking into account possible degeneracies), since the (spacelike) singularity is in the trapped region of the spacetime. For $d=2K+1$, the number of horizons depends on the value of the mass. For masses above $c_K$ the number is odd  and at least one horizon will always exist, whereas for masses below this critical mass the number of horizons will change to an even quantity, and will actually disappear at some point. 

The case where the EH-branch ends up at a maximum for some positive value of $g=g_\star$ ($r=r_\star$), or (b) type branch, is even simpler. There is a critical value of the mass for which $r_+=r_\star$. Below that mass a naked singularity appears. This is very similar to the situation described in the previous paragraph, the only difference being that in that case the radius of the singularity is zero, $r_\star=0$.

\begin{figure}[h]
\centering
\includegraphics[width=0.41\textwidth]{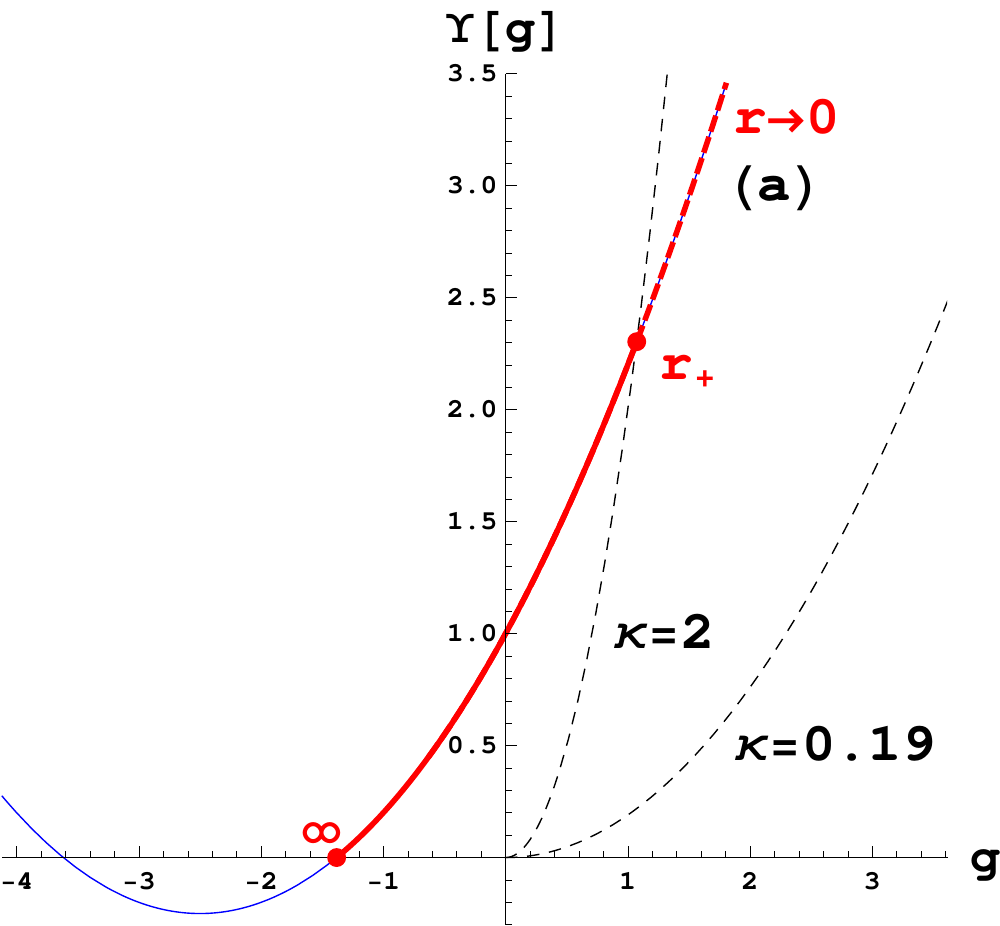} ~~\includegraphics[width=0.41\textwidth]{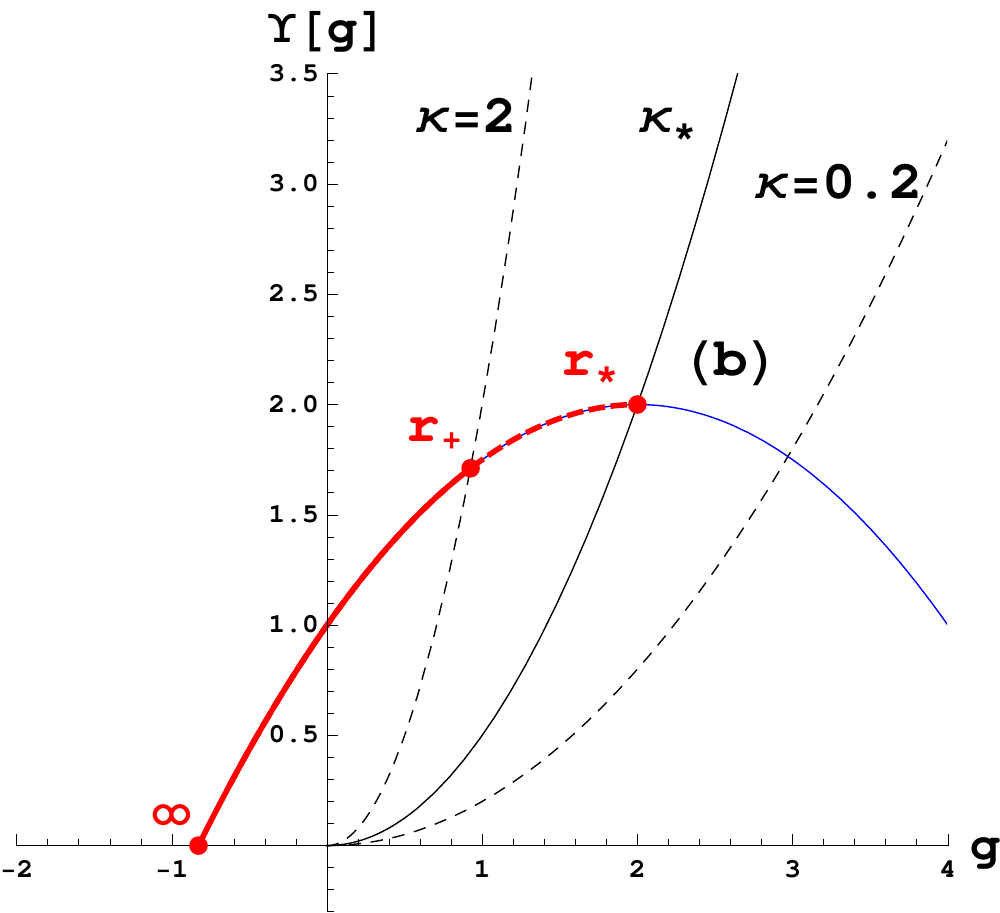} 
\caption{EH-branch in GB gravity  in 5 dimensions for $\lambda=1/5$ and $\lambda=-1/4$ respectively ($L=1$). The dashed curves correspond to the $\kappa\left( g/\sigma \right)^{\frac{d-1}{2}}$ for spherical topology ($\sigma=1$). The singularity becomes naked for $\kappa\leq \lambda=0.2$ in the first case and for $\kappa\leq \kappa_{\star}=0.5$ in the second case. For higher dimensions, the second figure would be qualitatively the same with just a different value of the critical mass. The first one, instead, changes as long as the horizon exists for all positive values of $\kappa$ in that case. For the (b) type branches the singularity is always located at $r_{\star}=\sqrt{-2\lambda}\,L$.}
\label{relevantGB}
\end{figure}

The simplest example of Lovelock theory is GB gravity \cite{Cai2002}, where we have just two branches, one of them suffering from BD instabilities. The remaining branch is thus an EH-branch. For $\lambda>0$ this branch is of the (a) type, extending all the way up to a singularity situated at $r=0$. For $\lambda<0$, instead, this branch has a maximum at positive values of $g$ (see figure \ref{relevantGB}).
This is a singularity at a finite value of $r$ that may or may not be naked depending on the value of the mass. The mass for which the horizon coincides with the singularity is
\begin{equation}
\kappa_{\star}=\frac{1}{2} (1-4\lambda) (-2\lambda L^2)^{\frac{d-3}{2}} ~.
\end{equation}
For bigger masses we have a well defined horizon while below this bound the singularity becomes naked.

Another intriguing possibility, that cannot be observed within the simple setting of GB gravity, is the would be appearance of several black hole horizons. For this to happen we need inflection points in $\Upsilon[g]$, and so the minimal example would be the cubic Lovelock theory. In the critical $d=7$ case, we have the possibility of obtaining two black hole horizons for some regions of the space of parameters, while this number is three in higher dimensions. One remarkable thing worth noticing here is that, as couples of horizons appear or disappear when we vary the value of the mass, the value of (the biggest horizon) $r_+$ may change discontinuously. Thus, the temperature as a function of the mass also varies discontinuously when crossing the values of $\kappa$ for which the outermost couple of horizons appear or disappear. One of the sides of the discontinuity has zero temperature (since the black hole is extremal for such critical mass) while the other has finite temperature. We must recall that the inner horizons are in general unstable \cite{Matzner1979,Poisson1990,Brady1995,Dafermos2003} and this possibly means that we should not trust our solution behind the outermost inner horizon. We may interpret these extremal states as black hole ground states, each one for a given range of masses. These seem naively accessible by evaporation and, thus, point towards a violation of the third law of black hole dynamics \cite{Torii2006}. They are in general unstable solutions \cite{Anderson1995a,Marolf2010}, though. At low temperatures, this particular branch will have several possible black hole masses and transitions might occur among them and the thermal vacuum \cite{Camanho2012b}. As for temperatures close to zero the free energy essentially coincides with the mass, the globally preferred solution is the less massive one: the vacuum. We will comment more on this later on.

\subsection{AdS (other than EH) branches}

The second class of branches that we describe in what follows are asymptotically AdS black holes different from the EH-branch. The latter will be included just for the discussion of negative mass solutions since the analysis is exactly the same. 

Consider first the positive mass solutions, for which the AdS branches always end at a maximum of the polynomial (see table \ref{AdS-case}).
\begin{table*}[h]
\centering
\begin{tabular}{c||c|c|c|}
asympt. & $\sigma=-1$ & $\sigma=0$ & $\sigma=1$ \\
\hline
\raisebox{9.3ex}{AdS}  & \includegraphics[width=0.26\textwidth]{AdSh2} & \includegraphics[width=0.26\textwidth]{AdSf} & \includegraphics[width=0.26\textwidth]{AdSs} \\
\hline
\end{tabular}
\caption{Taxonomy of the AdS (other than EH) branches.}
\label{AdS-case}
\end{table*}
As before, the existence of a horizon cloaking the singularity fixes the topology of such branches: As the considered sections of the polynomial run over negative values of $g$, horizons exist just for $\sigma=-1$. On the other cases, the solutions describe a spacetime with a timelike naked singularity. The condition for the existence of a horizon sets an upper bound on the mass, $\kappa<\kappa_{max}$, $\kappa_{max}$ corresponding to the critical value for which the radius of the horizon coincide with that of the singularity ($r_+=r_{max}$), {\it i.e.},
\begin{equation}
\kappa_{max}=r_{max}^{d-1}\;\Upsilon[g_{max}]~.
\end{equation}

When we encountered a naked singularity with positive mass in the EH-branch, it corresponded to the low mass limit of a multi-horizon black hole (below a given mass, the two horizons merge and disappear altogether, leaving the singularity naked). The case here is somehow different as the black hole horizon cannot degenerate as we increase the mass approaching the critical value. Thereby, the solution infinitesimally close to the critical one has non-zero temperature, diverging as we approach the bound. This is more reminiscent of the low mass limit of Schwarzschild black holes (ultimately leading to a regular geometry) than of the usual naked singularities in Einstein-Hilbert gravity. 

\begin{figure}[h]
\centering
 \includegraphics[width=0.55\textwidth]{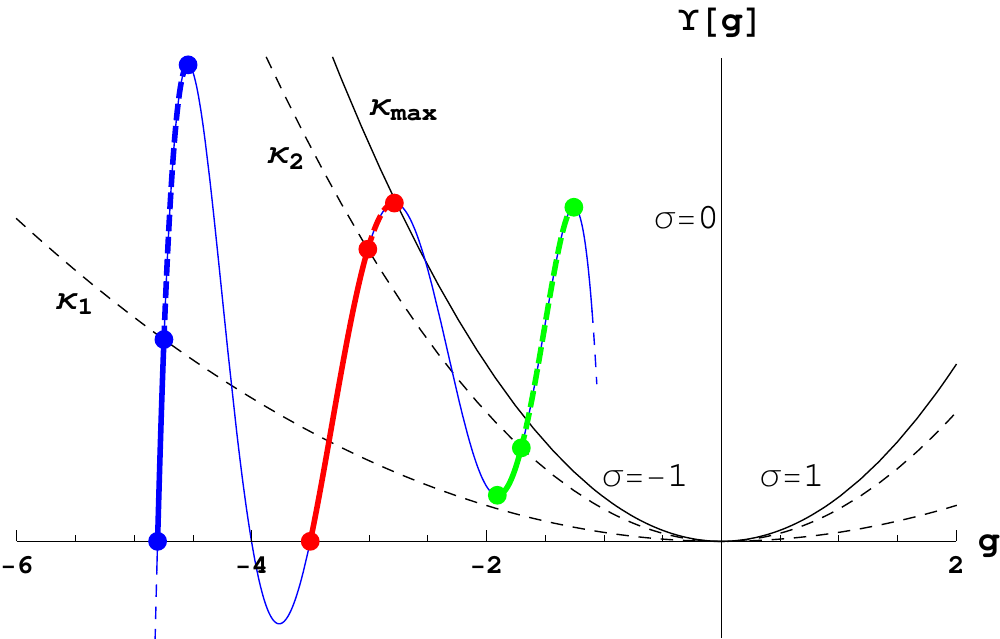}
\caption{Qualitative representation of (positive mass) AdS branches. The blue and red branches will be referred to as type (b1) and (b2) respectively, when considered for negative masses. The green one is an excluded branch. There are asymptotically AdS massive black holes for $\kappa<\kappa_{max}$ (red branch), otherwise the geometry displays a naked singularity.}
\label{horizonsAdS}
\end{figure}

For negative mass solutions, the analysis of the existence of black hole horizons and its number is more involved. The main qualitative feature is the possibility of having a minimum of the polynomial associated to the branch under analysis (see blue and red branches in figure \ref{horizonsAdS} for instance).
We will refer to the case without such minimum (blue branch) as type (b1) solution and as type (b2) for the other one (red branch). The structure of horizons and the type of singularity will differ in both types of branches. 

\begin{figure}[h]
\centering
\includegraphics[width=0.41\textwidth]{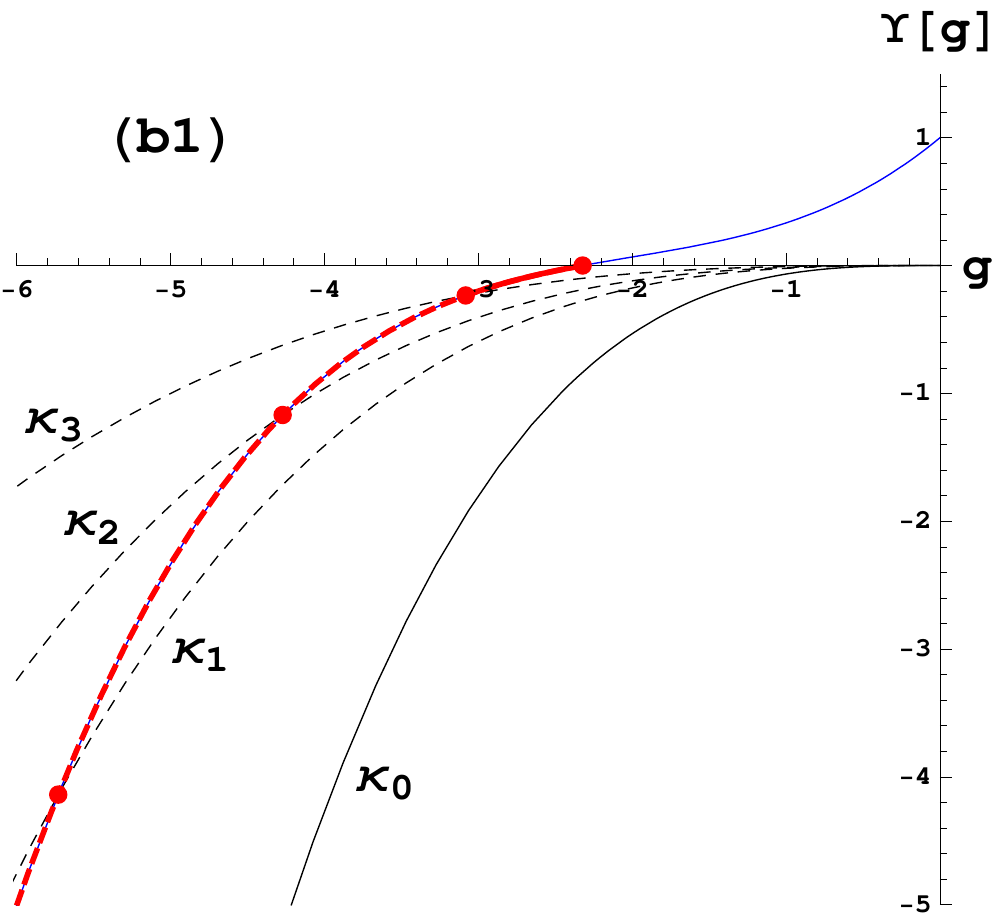} 
\caption{EH-branch in the cubic theory in 7 dimensions for $\lambda=0.4$ and $\mu=0.2$ ($L=1$). The dashed curves correspond to $\kappa\left( g_+/\sigma \right)^{\frac{d-1}{2}}$ with $\kappa_0=-\mu/3=-2/30$, $\kappa_1=-0.022$, $\kappa_2=\lambda=-0.015$, and $\kappa_3=-0.008$ ($\sigma=-1$). For masses above $\kappa_0$ we have one horizon with no distinction between positive and negative masses. For $\kappa \leq \kappa_0$ there is a naked singularity at $r=0$. No extremal state exists.}
\label{nohyperbolicground}
\end{figure}

When $d=2K+1$ and the branch we are considering is of (b1) type, there is a minimal mass for which, instead of an extremal regular spacetime with a degenerate horizon, we have a naked singularity (see figure \ref{nohyperbolicground}).
The temperature also vanishes asymptotically as we decrease the mass. We will not comment further on these kind of solutions as they are gravitationally unstable against perturbations \cite{Camanho2012}.  

\begin{figure}[h]
\centering
\includegraphics[width=0.43\textwidth]{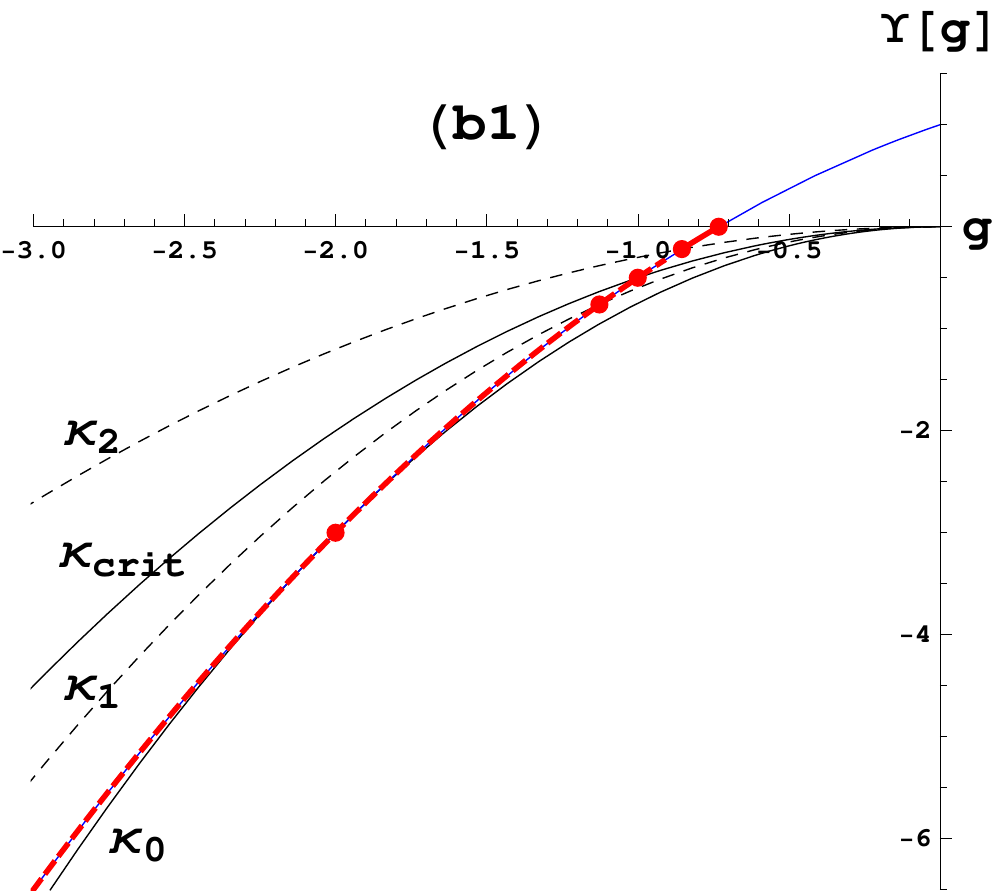} ~~\includegraphics[width=0.43\textwidth]{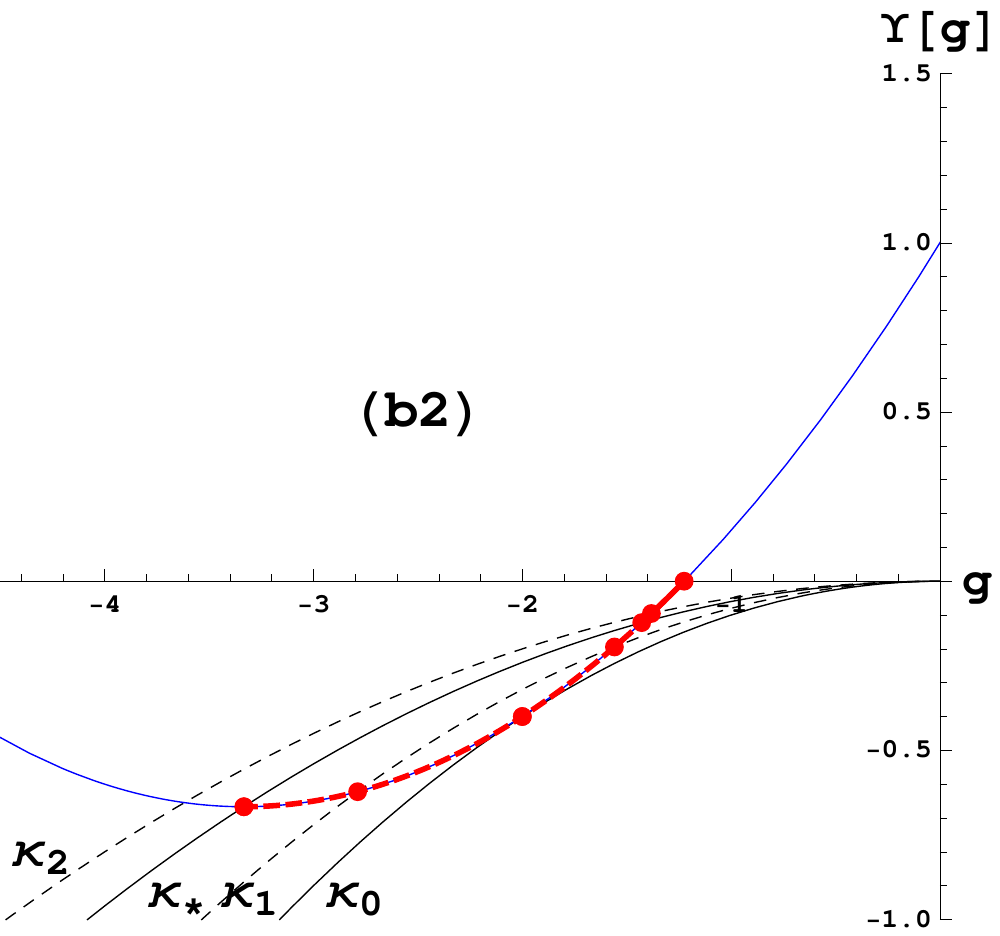} 
\caption{EH-branch in GB gravity  in 5 dimensions for $\lambda=-1/2$ and $\lambda=0.15$ respectively ($L=1$). The dashed curves correspond to $\kappa\left( g_+/\sigma \right)^{\frac{d-1}{2}}$ with $\kappa_0=-0.75$, $\kappa_1=-0.6$, $\kappa_{\rm crit}=\lambda=-0.5$, and $\kappa_2=-0.3$ (left) and $\kappa_0=-0.1$, $\kappa_1=-0.08$, $\kappa_\star=-0.06$ and $\kappa_2=-0.05$ (right) ($\sigma=-1$). In both cases we have two horizons for $\kappa_0<\kappa<\kappa_{\star}$ and one for $\kappa_\star<\kappa$. For $\kappa=\kappa_0$ we have a degenerate horizon. The singularity becomes naked for $\kappa\leq \lambda-1/4$ in both cases. In higher dimensions, the behavior is qualitatively the same. For the (b2) type branches the singularity is always located at $r_{\star}=L\sqrt{-2\lambda}$, while it is at the origin in the (b1) case.}
\label{negGB}
\end{figure}

In case we have a well defined extremal negative mass black hole, we always have at least two horizons for (b1) type branches and $d>2K+1$, as we depart from extremality. For (b2) type branches, however, the inner horizon disappears when its radius coincides with the radius of the singularity, changing its nature from timelike to spacelike, or viceversa. There is a critical mass for which this happens. This is irrelevant for an outside observer who cannot extract information from the inner horizon. Figure \ref{negGB} shows both kinds of solutions in the simplest case of GB gravity, for (b1) type ($\lambda<0$) and (b2) type ($\lambda>0$), respectively.   

Both classes of branches may have several horizons in the presence of inflection points. The (b1) type will always have an even number, except when $\kappa\geq c_K$ where the smallest horizon disappears. Thus, the singularity at $r=0$ is timelike below this critical mass and spacelike above it. The same change of behavior for the singularity appears in the other type of branches with the critical mass being set by the minimum, $\kappa_{\star}$. In the same way as described for spherical black holes in the EH-branch, the couples of horizons appearing or disappearing as a function of the mass translate into discontinuous changes on the temperature. The possibility of having several extremal solutions in one branch amounts to several ground states for different ranges of (negative) masses, with possible transitions among them. At zero temperature, though, one does not have a thermal vacuum to compare with, contrary to what happens at finite temperature. The negative mass black holes are the preferred phase in that regime.

\subsection{dS branches}

The existence of event horizons will again set a series of constraints both on the admitted topologies as well as in the possible values for the mass parameter. For hyperbolic or flat horizons, the metric function $f$ is always negative and the solution describes a big crunch. For $\sigma=1$ the solution may still describe a big crunch if we consider a high enough mass, above the Nariai mass. Slightly below it, at least two horizons exist, the biggest one being the cosmological horizon. Notice that there may be several Nariai masses.
\begin{table*}[h]
\centering
\begin{tabular}{c||c|c|c|}
asympt. & $\sigma=-1$ & $\sigma=0$ & $\sigma=1$ \\
\hline
\raisebox{9.3ex}{dS}  & \includegraphics[width=0.26\textwidth]{dSh} & \includegraphics[width=0.26\textwidth]{dSf} & \includegraphics[width=0.26\textwidth]{dSs} \\
\hline
\end{tabular}
\caption{Taxonomy of the dS branches.}
\label{dS-case}
\end{table*}

\begin{figure}[h]
\centering
 \includegraphics[width=0.53\textwidth]{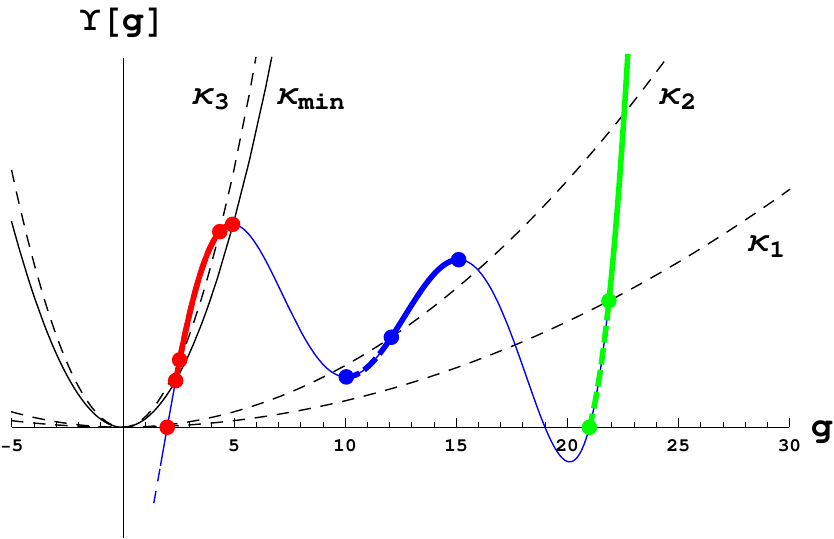} \qquad  \includegraphics[width=0.28\textwidth]{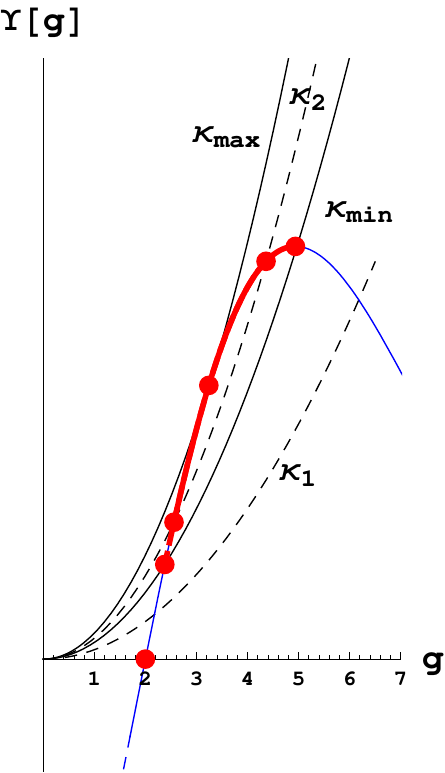}
\caption{Generic features of dS branches are captured in these figures. The different sections of the polynomial qualitatively represent (b) type (in red), excluded (in blue) and (a) type (in green) dS branches. The right figure zooms on the (b) type dS branch showing that for $\kappa_2$, $\kappa_{\rm min} < \kappa_2 < \kappa_{\rm max}$, there are asymptotically dS massive black holes, with outer and inner horizons. At $\kappa_{\max}$ the black hole becomes extremal. For $\kappa$ not in this range, the red branch displays a naked singularity. Even though the (a) type branch does not seem to have a black hole horizon (in the left figure, {\it e.g.}, the green dots correspond to the cosmological horizon), it always exists for $d>2K+1$ and, for some range of masses, also in the $2K+1$ dimensional case.}
\label{horizonsdS}
\end{figure}

Like the EH-branch, the dS branches can end up at a maximum, (b) type, or extend all the way to $r=0$ (a) type; red and green branches on figure \ref{horizonsdS} respectively.
For (b) type branches, there is a critical mass, $\kappa_{\rm min}$, for which the innermost black hole radius coincides with the radius of the singularity, $r_-=r_\star$. Below that mass, the horizon disappears. A naked singularity will always show up for sufficiently low masses.

For (a) type branches, the same happens if $d=2K+1$ just replacing the critical mass by $\kappa_{\star}\equiv c_K$. Again, we cannot avoid naked singularities as we approach arbitrarily low masses. Therefore, the existence of horizons sets, in this case, two bounds for the mass. Above the upper bound the geometry displays a spacelike singularity, whereas below the lower bound it has a naked timelike singularity. For $d>2K+1$, instead, the black hole horizon always exists, all the way down to zero mass. Thus, solely the upper Nariai bound for the mass exists.  

Another situation that we may encounter, for $d=2K+1$, is the absence of a Nariai solution. Below a certain critical mass we are faced with a naked singularity, the only remaining horizon being the cosmological one. This is the symmetric situation to the non-existence of extremal negative mass black hole discussed earlier for AdS branches. Mirroring that case, here we also find solutions with a large number of horizons. Their variation, as a function of $\kappa$, may translate into discontinuities in the temperature as a function of the mass. The discussion regarding how to interpret this phenomenon is the same as before.

\section{Heat capacity and local thermodynamic stability}

The details of the solutions and the behavior of their associated thermodynamic quantities strongly depend on the particular case under consideration. Therefore, a general thermodynamic analysis is cumbersome. However, once we have described the qualitative features of the different branches of solutions, much and very interesting information can be extracted, most of it arising as universal features of these black hole solutions.  

For positive mass black holes, the sign of the heat capacity on a BD stable branch, because of (\ref{dMdr}),  depends just on
\begin{equation}
\frac{dT}{dr_+}=-\frac{g_+}{2\pi}\left[(d-2)-\frac{d-1}{2}\frac{\Upsilon[g_+]}{g_+\Upsilon'[g_+]}\left(1+2g_+\frac{\Upsilon''[g_+]}{\Upsilon'[g_+]}\right)\right] ~,
\label{dTdr}
\end{equation}
where the second term (in brackets) seems to be related to the potential felt by perturbations in the shear channel \cite{Camanho2012}. We did not manage to check classical stability in full generality, but in the regimes of {\it high} and {\it low} masses. This should not be confused with the stability analysis focusing on perturbations of the black hole solutions (earlier relevant works on this include \cite{Dotti2005b,Gleiser2005b,Beroiz2007,Takahashi2009h,Takahashi2009d,Takahashi2010e,Takahashi2010d}).

We will just consider solutions possessing event horizons, since they are the only ones with associated thermodynamic variables. Then, taking into account our earlier discussion (see table \ref{cases}), we will have to deal with hyperbolic AdS branches; hyperbolic, flat and spherical EH-branches and spherical dS branches, classified in different subclasses. 

\subsection{Black holes in the EH-branch}

The simplest case to analyze is that of toroidal or planar black holes in the EH-branch. The thermodynamic variables, in this case, do not receive any correction from the higher curvature terms in the action and the expression reduces to the usual formula of Einstein-Hilbert gravity,
\begin{equation}
\frac{dT}{dr_+}=\frac{d-1}{4\pi L^2} ~.
\end{equation}
This expression is positive. Therefore, these black holes are locally thermodynamically stable for all values of the mass. 

For the EH-branch, the only one admitting all three topologies, the situation is exactly the same for $\sigma=\pm 1$ in the large mass limit. The value of $g$ at the horizon, $g_+=\sigma/r_+^2$, asymptotically approaches zero from both sides, and the formulas reproduce the planar case. Einstein-Hilbert gravity captures the universal thermodynamic description of Lovelock black holes with large enough mass. It does not capture other features in this regime. For instance, those related to the stability and causality preserving properties of the solutions \cite{Camanho2010a,Boer2009a,Camanho2010d}. EH-branch's black holes in this regime are then always stable.

This is also the case for a special class of (maximally degenerated) Lovelock theories whose analysis is not considered in the present article. Those theories admit a single (EH-)branch of black holes \cite{Crisostomo2000}. The results there also coincide with the generic analysis for spherical black holes presented here.

It is worth recalling here that for sufficiently large mass, just the EH-branch admits black holes, the other branches describing geometries with naked singularities or big crunch spacetimes.

\begin{figure}[h]
\centering
\includegraphics[width=0.46\textwidth]{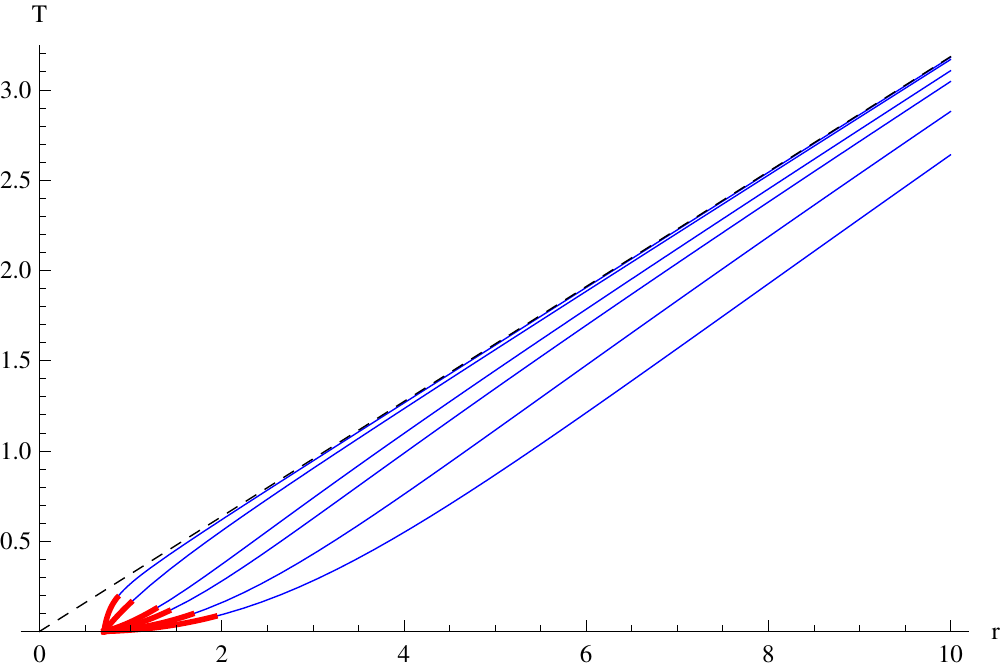}\qquad \includegraphics[width=0.46\textwidth]{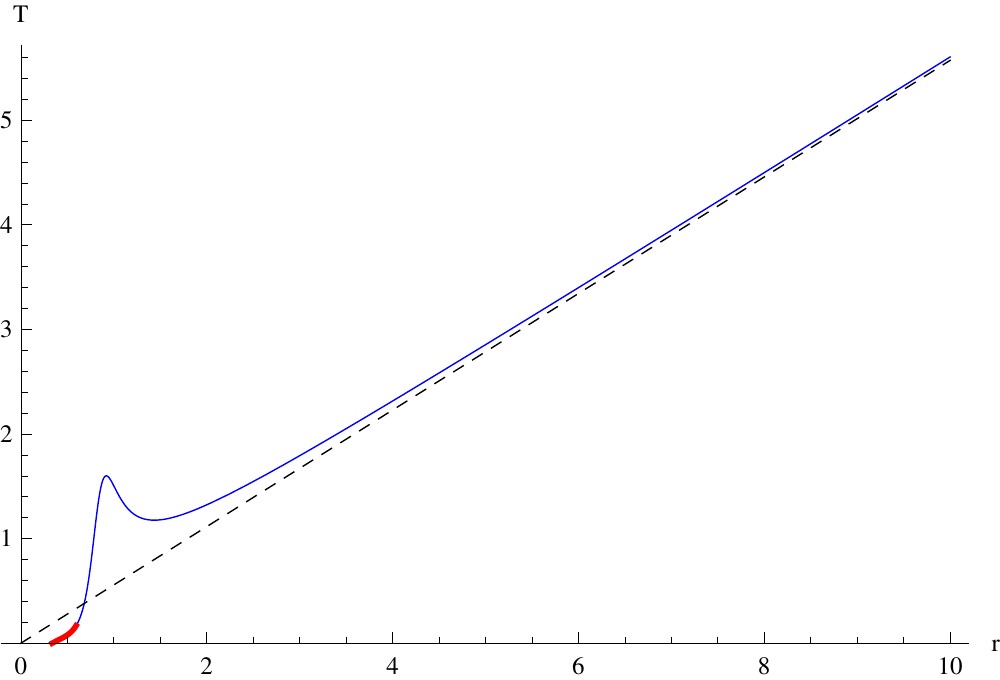} 
\caption{Temperature versus horizon radius (equivalently versus black hole mass, $\kappa$) for hyperbolic black holes ($L=1$). The first figure corresponds to GB gravity in $d=5$ for $\lambda=-10,-5,-2,-1,0,0.2$ (from bottom to top). The behavior is qualitatively the same in higher dimensions. Such black holes are stable for all values of the mass (even negative, in red). The black dashed line corresponds to the planar case to wich all curves asymptote. The second figure corresponds to an example of intermediate unstable phase in cubic Lovelock theory in $d=8$ with $\lambda=0.65$ and $\mu=0.5$.}
\label{stabGBh}
\end{figure}

As we decrease the mass, particular features of the different topologies pop up and we need to consider them separately. Small mass hyperbolic black holes correspond to a smooth deformation of the vacuum. In that regime, the second term inside the brackets of (\ref{dTdr}) becomes negligible, $\Upsilon[g_+] \approx \Upsilon[\Lambda] = 0$, and the expression approaches
\begin{equation}
\frac{dT}{dr_+}\approx \frac{d-2}{2\pi}\,(-g_+) ~.
\end{equation}
Therefore, as $g_+<0$, the low mass hyperbolic black holes are also stable. Notice, though, that even if both extrema of the spectrum for hyperbolic black holes in the EH-branch are stable, one may encounter unstable intermediate phases. This is not the case in GB gravity (contrary to what is stated in \cite{Neupane2004}; the negative specific heat found there corresponds to inner horizons and, thus, does not indicate any instability of the system), where we find that hyperbolic black holes are always locally thermodynamically stable, as can be seen in figure \ref{stabGBh} (left), even for negative masses above the extremal one (depicted in red in the figure).
The cubic theory, in turn, as shown in figure \ref{stabGBh} (right), already displays intermediate mass hyperbolic black holes which are locally thermodynamically unstable.

The extremal hyperbolic black hole can be shown to be always stable from this point of view. It has zero temperature and all black holes with higher masses have positive temperature. Thereby, the heat capacity has to be positive, close enough to this state. In some cases, for $d=2K+1$, the extremal negative mass black hole does not exist. In that situation we may consider, in principle, infinitesimally small black holes, $r_+\rightarrow 0$, with temperatures approaching zero asymptotically. The {\it singular} `zero size black hole', again, fixes a bound in the mass, and black holes close to that bound are stable, in the same way as the ones close to the extremal black hole.

For spherical black holes we have to distinguish between different cases. In the (a) type, the small black hole limit, $r_+\rightarrow0$, may correspond respectively to finite or zero mass for $d=2K+1$ and $d>2K+1$. If the branch is of (b) type, there is a lower bound for the mass of the black hole for which the temperature diverges. When $d>2K+1$, in the low mass regime,
\begin{equation}
\frac{dT}{dr_+}\approx -\frac{d-2K-1}{4\pi\,K}\,g_+ < 0 ~.
\end{equation}
Thereby, small spherical black holes in the EH-branch are thermodynamically unstable, in exactly the same way as in the usual Einstein-Hilbert gravity. The situation changes for $d=2K+1$, where the temperature vanishes asymptotically. Therefore, small black holes are stable in odd dimension for the highest order Lovelock gravity (in particular, we need $c_K>0$ for type (a) EH or dS-branches to exist), whereas they are unstable in all other cases \cite{Cai2004}.

\begin{figure}[h]
\centering
\includegraphics[width=0.45\textwidth]{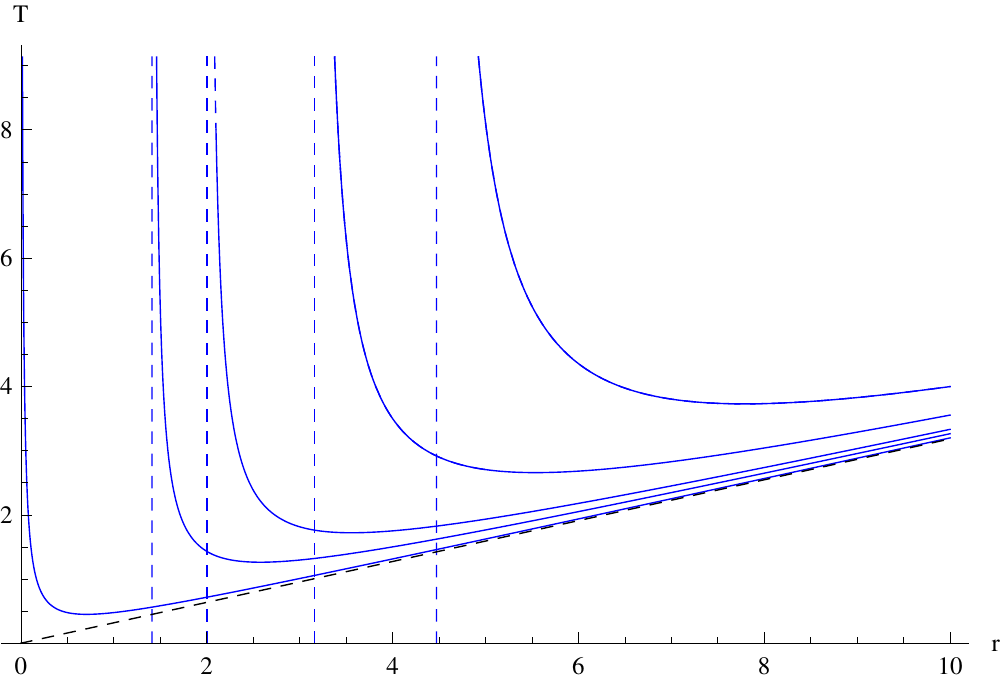}\qquad \includegraphics[width=0.45\textwidth]{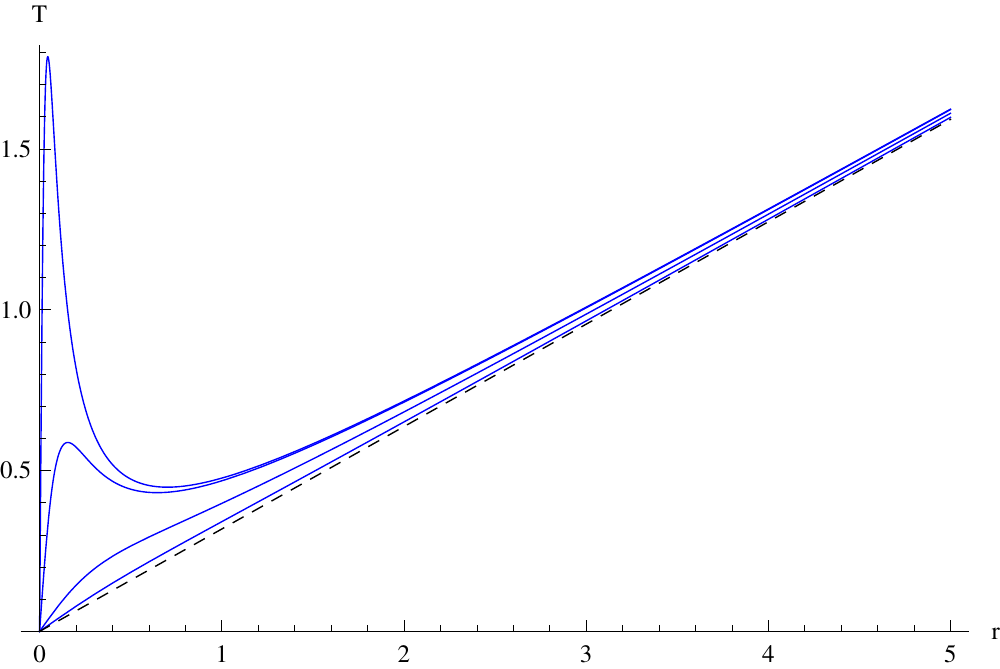} 
\caption{Temperature versus horizon radius (equivalently, mass) for spherical black holes in $d=5$ GB gravity ($L=1$). The first figure corresponds to negative values of the GB coupling, $\lambda=-10,-5,-2,-1,0$ (from top to bottom), whereas the second considers positive values, $\lambda=0.001,0.01,0.1,0.2$ (from top to bottom). The dashed blue lines indicate the value of the mass, $\kappa_{\rm min}$, for which the temperature diverges. We observe the appearance of a new stable phase for small positive values of $\lambda$ and even the disappearance of the unstable region for high enough $\lambda$. This stable region of small black holes disappears in higher dimensions, for all positive values of $\lambda$, the qualitative behavior being similar to the $\lambda=0$ case.}
\label{stabGBs}
\end{figure}

For (b) type branches something similar happens as we approach the minimal mass, $\kappa_{\rm min}$, set by the local maximum of the polynomial $\Upsilon[g]$. At this critical mass the temperature diverges, but in this case as we lower the mass. The heat capacity is negative, as we can also infer from (\ref{Upsmax}), taking into account that $g_+$ has a positive value. Black holes close enough to the minimal mass one are then unstable. This kind of behavior can be seen for instance in GB gravity (see figure \ref{stabGBs}).  

\begin{figure}[h]
\centering
\includegraphics[width=0.47\textwidth]{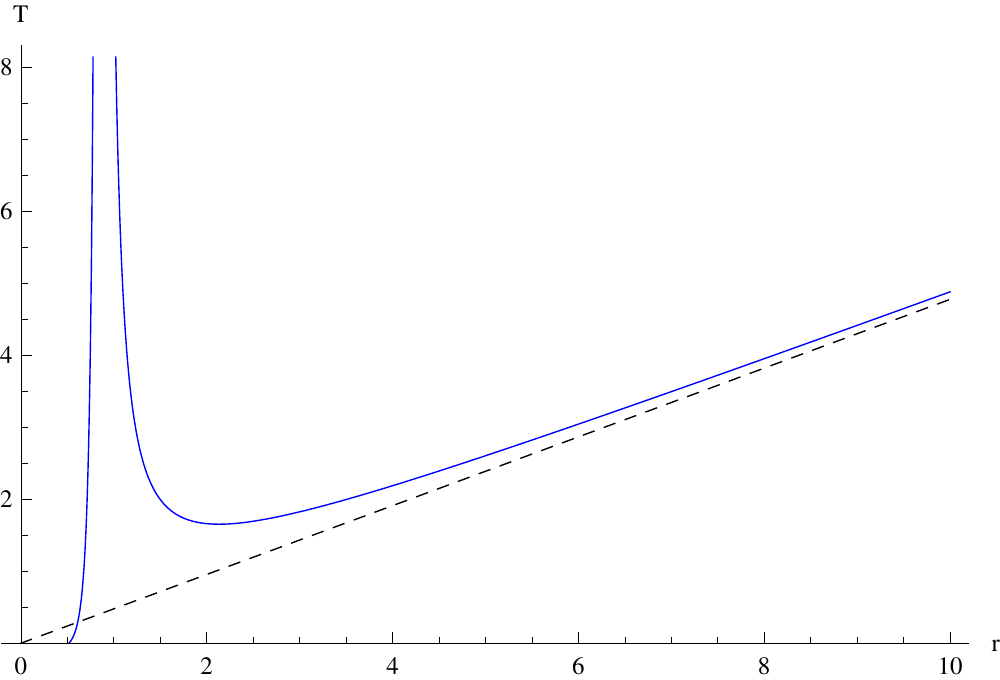}~~ \includegraphics[width=0.47\textwidth]{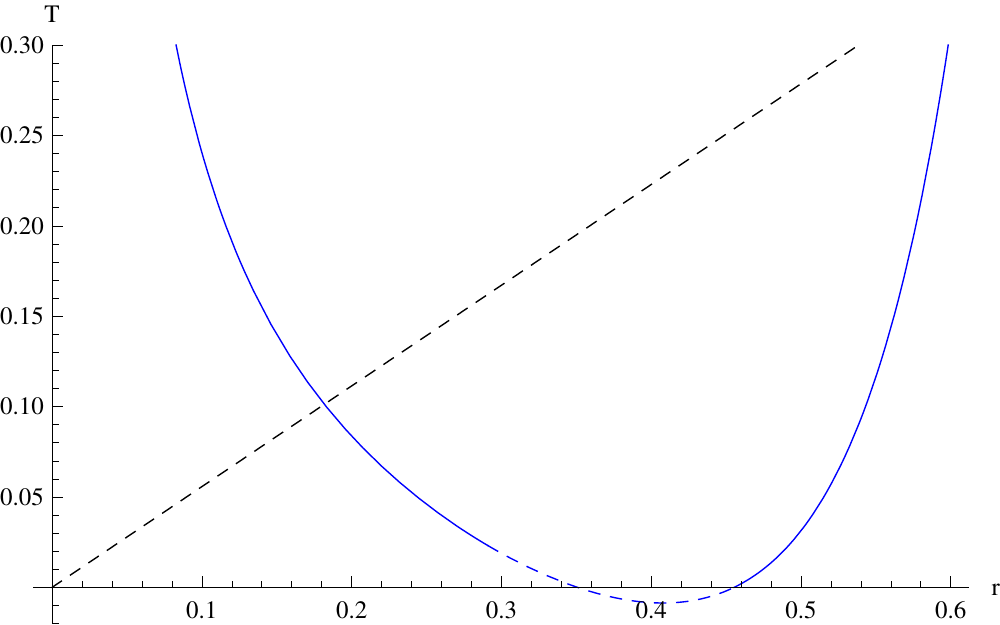} 
\caption{Temperature versus horizon radius (equivalently, mass) for spherical black holes in $d=7$ cubic Lovelock gravity with $\lambda=-0.746$, $\mu=0.56$ and $L=1$ (left). The black hole reaches zero temperature for finite (positive) mass. For $d=8$ (right), the shape of the curve is qualitatively the same except in the low radius region where a new branch of black holes with diverging temperature appears. We just show a zoom of the left bottom corner. We can also verify the existence of the temperature (and radius) jump commented in the main text. The dashed blue line corresponds to inner horizons, one of them becoming outer for masses below the extremal one.}
\label{stable}
\end{figure}

There is one further possibility when the polynomial is such that it allows for more than one black hole horizon. Then, either for (b) type as well as (a) type solutions (in $d=2K+1$ dimensions), we may still have an event horizon cloaking the singularity for some range of masses below the naive $\kappa_{\rm min}$. As we lower the mass further we encounter at least one extremal black hole, which is stable in the same way as the extremal hyperbolic black hole of negative mass.  One such case is shown in figure \ref{stable}, that corresponds to the cubic polynomial plotted in figure \ref{horizons}.
In general, we may also have unstable regions in between the two stable ones, and also for masses below the extremal one if $d > 2K + 1$ (see figure \ref{stable}, right). This situation does not decisively depend on the dimensionality of spacetime. This case has features of the previously discussed spherical black holes but its behavior is similar to the (negative mass) near extremal hyperbolic black holes.

\subsection{Hyperbolic black holes in the AdS-branches}

Most of the discussion on hyperbolic black holes in the EH-branch also applies, on general grounds, to the AdS-branches. The only difference being that, in general, there is a maximal mass for which the temperature diverges. Thus, close enough to that point the heat capacity has necessarily to be positive and the black hole thermodynamically stable. This can be seen directly from (\ref{dTdr}), as the heat capacity close to the maximum approaches
\begin{equation}
\frac{dT}{dr_+}\approx\frac{d-1}{2\pi}\,\frac{g_+ \Upsilon[g_+]\,\Upsilon''[g_+]}{\Upsilon'[g_+]^2} ~,
\label{Upsmax}
\end{equation}
diverging as well when we reach the critical mass. $\Upsilon[g_+]$ is positive due to the positivity of the mass. The second derivative $\Upsilon''[g_+]$ is negative as we are close to a maximum, but $g_+$ is negative as well. The plot of temperature versus horizon radius will be in general qualitatively similar to that corresponding to the EH-branch, with the difference that the temperature diverges at some finite value of $r_+$. 

\subsection{Spherical black holes in the dS-branches}

The low mass regime of the spherical solutions corresponding to dS branches is very similar to that of the EH-branch. The high mass regime, instead, is very different. These black holes may increase their mass until they reach a maximal (so-called {\it Nariai}) mass, which is set by the shape of the polynomial. This is an extremal state with zero temperature and, as we reach it from lower mass configurations, the system is thermodynamically unstable close to it.

\begin{figure}
\centering
 \includegraphics[width=0.47\textwidth]{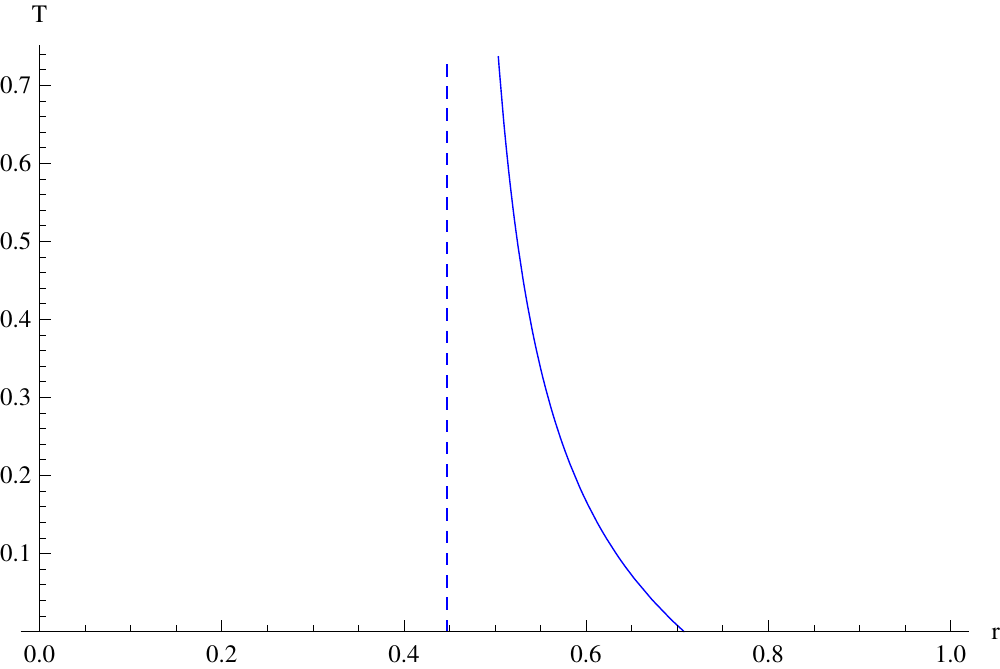}\quad \includegraphics[width=0.47\textwidth]{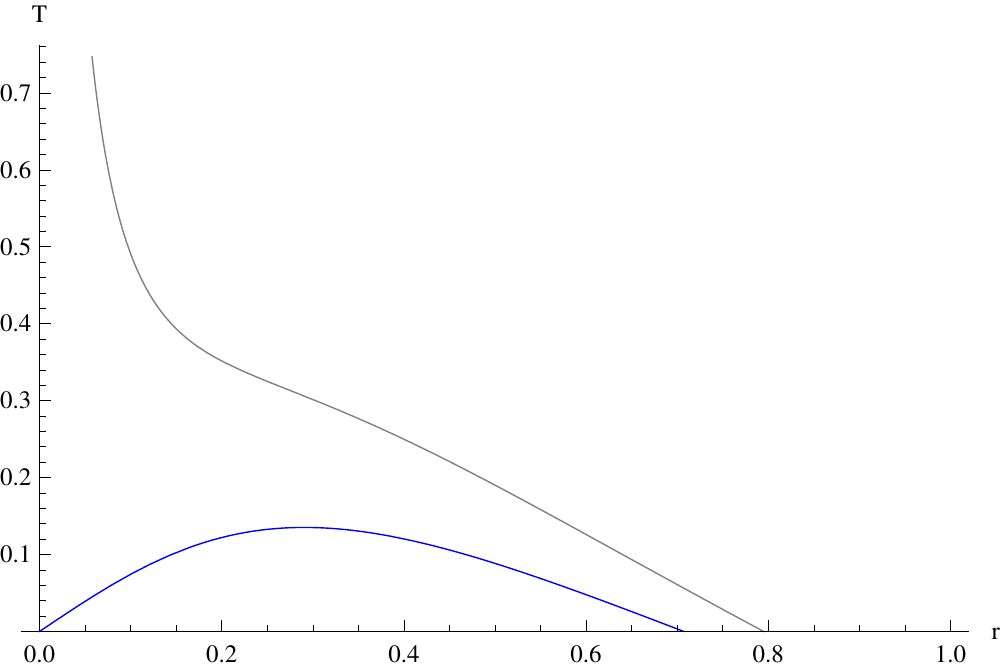} 
\caption{Temperature versus horizon radius (equivalently, mass) for spherical black holes in $d=5$ GB gravity with positive cosmological constant, $L^2=-1$ (in blue). The first figure, $\lambda=-0.1$, corresponds to a (b) type dS branch. We can identify the zero temperature state in the high mass regime with the Nariai solution. The temperature diverges as we approach the lower bound, $\kappa_{\rm min}$, indicated by the dashed blue line. For $\lambda=0.1$, in the second figure, a spherical (a) type dS branch arises. We may identify again the extremal state with maximal mass with the Nariai solution, for which the temperature goes to zero, as well as the radius of the black hole horizon. The `zero size black hole' has finite mass in this case. For higher dimensions the stable region of small black holes disappears (gray line corresponds to $d=6$), the spherical black holes being unstable as their (b) type counterparts. The only difference is that the temperature diverges in the zero mass limit.}
\label{dSbranch}
\end{figure}

We may construct dS branches of (a) and (b) types using GB gravity with positive cosmological constant (setting $c_0=-L^{-2}$). In this case, the (b) type branch (figure \ref{dSbranch}, left) is unstable for all allowed values of the mass, since we are considering $d=5$ (which is $d=2K+1$ in this case), while a stable region of small black holes appears for (a) type branches (figure \ref{dSbranch}, right) \cite{Cai2004a}.
This stable region disappears in higher dimensions. In general, spherical dS branches may have some stable intermediate region, but they are unstable (or even non-existent) for high enough temperature.

Let us summarize the main results of this section. Hyperbolic black holes generically have two stable domains, at low and high temperatures. For intermediate temperatures, these solutions may have more than one possible mass, some of them unstable. The only difference of AdS branches with respect to the EH case is that the high temperature regime has a maximal finite mass in the former. 

The case of spherical black holes exhibit quite distinct features. From the thermodynamical point of view, we may distinguish those situations where there is, or there is not, a minimal temperature, $T_{\rm min}$, for the black holes to exist. In the former case, we do not reach any extremal black hole, neither in the low mass, nor in the high mass regimes. In the case of the EH-branch, for $T < T_{\rm min}$, only the thermal vacuum may exist whereas, for high enough temperatures, a black hole may exist with two very different masses, one close to $\kappa_{\rm min}$ (which is unstable) and one very high (stable). For intermediate temperatures close to $T_{\rm min}$, we may in principle encounter several stable and/or unstable black holes. In the latter case, instead, black hole solutions exist for the whole range of temperatures, except in the case where a dS-branch reaches an extremal state at low as well as at high masses. For the EH-branch, in this situation, we have two stable phases again, one in the low and one in the high temperature regimes. At intermediate temperatures, the black hole may have several possible masses, some of them unstable. 
 
\section{Hawking-Page-like phase transitions}

The existence of unstable phases, as well as the several possible black hole solutions at the same temperature, suggest the occurrence of Hawking-Page-like phase transitions, as already observed in the case of Gauss-Bonnet gravity \cite{Cvetic2002,Nojiri2001j}. These phase transitions should be particularly relevant when studying the physics of the dual CFT plasma. In order to analyze this we need to discuss the global stability of the solutions. Any system in thermal equilibrium with an infinite heat reservoir (and thus at constant temperature) will be described by the canonical ensemble, whose relevant thermodynamic potential is the Helmholtz free energy, $F$. The preferred, and so globally stable, solution is the one that minimizes $F$. For instance, the free energy of the black hole solution calculated in (\ref{free-energy}), is the free energy with respect to the vacuum solution, except for the hyperbolic case where the finite ground state free energy must be subtracted. Therefore, the sign of the free energy determines which solution is globally preferred at any given temperature, the appropriate black hole (if several are possible) or a thermal bath in vacuum.

The general analysis is, again, hard and not very enlightening. We will just concentrate in showing general features of these black hole solutions without entering into the details of the different cases. We will consider the same regimes analyzed for the local stability, as there we can easily find the expression for the free energy. In the planar case the analysis is very simple since the free energy reads
\begin{equation}
F=-\frac{V_{d-2}}{16\pi G}\frac{r_+^{d-1}}{L^2} ~.
\end{equation}
The black hole is then always the preferred solution, as indicated by the negative sign of the free energy, and no phase transitions occur. This will be the situation in the large mass limit of the other topologies in the EH-branch. As for large enough $r_+$ the free energy may be as large as one wants, ambiguities on the reference background do not matter in this limit.   

\subsection{Spherical black holes}

For spherical black holes, we will restrict our discussion to the two most generic situations. For the EH-branch, we will consider separately the case of having a stable low temperature phase (as in $d=5$ GB gravity with positive $\lambda$), and the case where a minimal temperature is needed for black hole solutions to exist. The second case is the analog of the usual situation in Einstein-Hilbert gravity. At low temperatures, the thermal vacuum can be considered as the globally stable solution whereas, for higher temperatures, two or more black hole solutions are possible. For high enough temperature just two of them remain. The small one has always positive free energy. For an (a) type branch,
\begin{equation}
F=\frac{(d-2)V_{d-2}}{16\pi G}\frac{r_+^{d-2K-1}}{d-2K} ~,
\label{freesmall}
\end{equation}
whereas for the (b) type case the temperature diverges as we approach the maximum $g_+\rightarrow g_\star$, and the free energy is $F\approx -T S$. Then, for positive entropy, the small black hole solution has negative free energy and is stable against the vacuum. However, it is not the minimum of the free energy since the big black hole has always a lower one. This is quite easy to see by realizing that the small black hole entropy goes to a constant as we approach $g_\star$ whereas the entropy grows indefinitely for big black holes, since they approach the planar limit. Thus, the small black holes are not just locally but also globally unstable. 

The big black holes have, in general, negative free energy. We have then a Hawking-Page-like phase transition, from the thermal vacuum at low temperatures to big black holes at high temperatures. The difference with respect to the Einstein-Hilbert case is that we may have several black holes at intermediate temperatures, with either sign of the free energy. Recall that, as we saw in (\ref{free-energy}), $F$ may have up to $2K-1$ zeros, that is, it can change sign those many times. Being an odd number, $F$ has different sign for small and big black holes. For ranges of temperature where several black holes have negative free energy, transitions among them may happen, the globally preferred solution being the one with the lowest free energy. This would be an example of a new kind of phase transition, different from the Hawking-Page one, where one of the phases is always the thermal vacuum. Further research on this possibility is currently under scrutiny \cite{Camanho2012b}.

If the EH-branch has stable low temperature black holes, {\it i.e.}, for (a) type in $d=2K+1$, these are globally unstable as indicated by their positive free energy that asymptotes a constant when $r_+ \to 0$,
\begin{equation}
F=\frac{(d-2)V_{d-2}}{16\pi G}~.
\end{equation}
This is very similar to the earlier formula (\ref{freesmall}). The same happens for the would be extremal black holes that one may encounter in the EH-branch. In the limit of low temperatures, the free energy coincides with the mass and, as such states have positive mass, they are globally unstable. The globally preferred phase is the thermal vacuum which is the minimal mass solution. Then, again, one has the same kind of transition described in the previous paragraph.

Another situation we did not comment at length is the possibility of having negative entropy for the spherical black hole with critical mass, $\kappa_{\rm min}$. This happens already in the simplest possible case of GB gravity, for negative $\lambda$, where there is a maximum in the EH-branch situated at $g_{\star}=-1/2\lambda$. As pathological as it may seem, the consequence of this from the global stability point of view is clear. Again, the globally preferred solution is the big black hole as before, and the discussion goes through. This is quite general: a negative entropy state necessarily has bigger free energy than the vacuum (characterized by minimal mass and vanishing entropy).

For dS branches the situation at low mass is exactly the same as for spherical solutions in the EH-branch. At low temperatures the free energy approaches the value of the mass and the globally stable phase is always thermal vacuum. For (a) type branches and $d >2K+1$, no high temperature black hole exists, thereby the preferred phase in that regime would be trivially the thermal vacuum and no Hawking-Page-like phase transition seems to occur (see, for instance, \cite{Cai2004a}, for the GB case). In any other situation ({\it e.g.}, (b) type branches) with positive entropy, the globally stable solution would be the near-critical black hole approaching the maximum of the polynomial. Therefore, these branches seem to display phase transitions, even though high temperature black holes are locally thermodynamically unstable.

\subsection{Hyperbolic black holes}

For the hyperbolic black holes one may compute the free energy at the high and low temperature regimes as before. As the maximally symmetric space has temperature in this case it is not clear how to use it as a ground state. Instead, we will consider the extremal negative mass black hole as the reference state --with vanishing entropy-- given that it can be identified with any temperature \cite{Hawking1995}\footnote{This has been disputed by some authors (see \cite{Anderson1995a} for instance) in reason of the semiclassical instability of these solutions.}, since the Euclidean section is regular for arbitrary period in imaginary time. Otherwise, the analysis would become trivial with just one or more black hole solutions, no matter the value of the temperature. No Hawking-Page-like phase transitions would occur in that case, just the possibility of transitions among black holes of different masses at intermediate temperatures.

For the EH-branch in the high temperature regime we just have one possible black hole solution. It has negative free energy as it approaches the planar limit and so it is globally preferred. The same happens for the AdS branches, that end up at a maximum of the polynomial. As we approach the critical mass, $\kappa_{\rm min}$, the free energy approaches $F\approx-TS$ that is arbitrarily negative for positive entropy. For negative entropy at the maximum, which corresponds to the biggest possible black hole in the AdS branch, every black hole has negative entropy. In this case the globally preferred phase is the reference state, since it has zero entropy and minimal mass. 

For lower temperatures we have to consider black holes close to the extremal one. In the zero temperature limit only these extremal states matter and their free energy is simply given by their mass. Then, the globally preferred phase in that limit is the lowest mass state. For slightly higher temperatures one expects that the globally preferred solution is still described by the same minimal mass extremal state (identified with finite temperature) or the corresponding black hole solution that is just a smooth deformation of it. It is however hard to elucidate in general which of both solutions has the lowest free energy, and then the existence or not of Hawking-Page-like phase transitions.

\section{Discussion}

In this paper we presented a novel approach to deal with the full classification and description of black hole solutions with constant curvature horizons in Lovelock gravity. Our proposal allows to treat the generic case where the whole set of Lovelock coupling constants is arbitrary, contrary to most studies existing in the literature where the analysis is restricted to particular cases. Most of these cases, moreover, correspond to degenerate vacua of Lovelock gravity, while our approach is valid in general and is most useful in the non-degenerate case.

We discussed the main features of all possible configurations. In particular, we have established a recipe to scrutinize the number of horizons and their evolution with the mass, something that we expect to be useful to visualize and gain intuition in physical processes involving black holes in such theories: evaporation, mass accretion and appearance of naked singularities \cite{Camanho2012}.

We presented some general features of Lovelock black holes' thermodynamics, analyzed their local and global stabilities and the possible existence of phase transitions. Even if these solutions show some seemingly pathological features, such as negative values for the entropy, these are avoided if we restrict ourselves to the globally preferred phase.

For asymptotically AdS solutions (either in the EH or AdS-branches), global stability in the high temperature regime always selects the biggest black hole, being the one with biggest entropy. If we apply the same criterium to all possible solutions, regardless of the branch to which they belong, the selected solution is always the one approaching the planar limit, since it is the only one that has arbitrarily big entropy. The comparison of solutions belonging to different branches is not really allowed as they have different asymptotics. The usual euclidean prescription says that we must compare all solutions {\it with the same boundary conditions} what certainly includes the asymptotics. However the existence of spherical shock wave solutions \cite{Gravanis2010a} separating regions corresponding to different vacua suggest the possible existence of mixed solutions and transitions between branches. In that context, the high temperature phase would always correspond to the universal planar limit \cite{Camanho2012b,Camanho2012c}. From a holographic point of view, this is consistent with the intuitive idea that effects due to the curvature of spacetime, being overcome by finite temperature effects, are negligible for field theory calculations. This is a new way of addressing the multi-valuedness problem of Lovelock theories.

The usual Einstein-Hilbert gravity admits in principle topological solutions displaying naked singularities. These may arise as a result of a bad choice of topology or as associated to negative mass, below the extremal one for hyperbolic horizons. The latter are just a special case of {\it trans-extremal} solutions, where the values of the parameters are chosen in such a way that, an otherwise well defined black hole with positive temperature, is beyond the extremal state. Example of this are the Reissner-Nordstrom or the negative mass hyperbolic black holes. In principle, all these situations are ruled out by the cosmic censorship conjecture that states that naked singularities do not form in the evolution of generic initial conditions. For instance, the evaporation process for black holes with several horizons should stop at the extremal state as it has zero temperature and this avoids the formation of trans-extremal solutions in that case.

The situation in generic Lovelock theories of gravity is rather different. For this wide family, there are several situations that suggest a possible violation of the cosmic censorship conjecture, as we have seen analyzing the case of static uncharged black holes. In addition to the cases pointed out in the previous paragraph, new kinds of naked singularities appear, some of them which naively seem to be formed in the evolution of these black holes. This can happen for the otherwise well-behaved Einstein-Hilbert branch, and it certainly happens generically on the extra `higher order' (A)dS-branches. Most of these naked singularities arise because the branch of interest ends up at a maximum or a positive (for positive mass) minimum of the Lovelock polynomial. The latter corresponds to a complex cosmological constant associated with this particular branch. The former, in turn, appear in a variety of cases. They constitute a maximal mass for hyperbolic black holes of AdS-branches. They also provide a minimal mass for some spherical black holes in the EH or dS-branches.

The other possibility for naked singularities to appear is just the spherical case in the maximal $d=2K+1$ Lovelock theory, for the EH or dS-branches when they extend all the way to $r=0$ without encountering any singularity (maxima or minima of $\Upsilon[g]$). In those cases we find a naked singularity for masses below a critical value. Any other possible naked singularity may be considered in the same class as those appearing in Einstein-Hilbert gravity. 

As we think of the evolution of the black holes studied in the present paper, we realize that naked singularities seem easy to form, at least naively. Consider for instance the evaporation of spherical black holes. For (b) type EH or dS branches these black holes always reach a critical mass where the horizon coincides with the singularity. At that point the temperature diverges but a finite mass naked singularity remains. The naked singularity inevitably forms. We emphasized the word {\it naively} before since the present analysis just considers the thermodynamic stability of the solutions. These solutions are locally and globally unstable, however they are still valid solutions that may form under evolution of generic spherically symmetric initial conditions. A more general analysis is needed to elucidate whether naked singularities may form or not in these theories \cite{Camanho2012}.

We have fixed, throughout this paper, the values of the cosmological constant and the Newton constant appearing in the lagrangian to their customary values in AdS Lovelock gravities. It is worthwhile mentioning that a straightforward generalization of this work amounts to studying the case of dS Lovelock theories (note that there are AdS vacua also in this case), as well as theories where the Newton constant has negative sign. On the one hand, we shall mention that this sign flip was already considered in the context of three dimensional topologically massive gravity, where it was found that a negative Newton constant is useful to render otherwise negative energy modes harmless for the stability about flat space \cite{Deser1982a}. Furthermore, in higher dimensions, even though $G_N < 0$, the generic structure of branches discussed in this paper will remain, and there will always be solutions corresponding to well-defined gravities with positive cosmological constant.

In the last few months there were some papers constructing gravitational theories that share some compelling properties with Lovelock lagrangians \cite{Oliva2010, Oliva2010a}. In particular, these are lower dimensional theories displaying black hole solutions whose profile precisely correspond to Lovelock black holes \cite{Myers2010b}. Some of the results of this paper are therefore of direct application to those cases as well. This is particularly interesting due to the fact that quasi-topological gravities are higher curvature theories in dimensions lower than their corresponding Lovelock cousins, thus the results are of interest in more `physical' setups of AdS/CFT \cite{Myers2010d}.

Lovelock theories have the remarkable feature that lots of physically relevant information is encoded in the characteristic polynomial $\Upsilon[g]$. Boulware-Deser-like instabilities, for instance, can be simply written as $\Upsilon'[\Lambda] < 0$, which has a beautiful CFT counterpart telling us that the central charge, $C_T$, has to be positive. Now, $\Upsilon'[\Lambda]$ can be thought of as the asymptotic value of the quantity $\Upsilon'[g]$ that is meaningful in the interior of the geometry, and has to be positive all along the corresponding branch. Naked singularities taking place at extremal points of the polynomial are suggestive of the fact that $\Upsilon'[g]$ should be a meaningful entry of the holographic dictionary (see \cite{Paulos2011} for related ideas) that does not exist in the case of Einstein-Hilbert gravity.

We hope to report on some of these interesting open problems elsewhere.

\ack
%
We would like to thank Andr\'es Gomberoff, Hideki Maeda, Julio Oliva, Miguel Paulos, Ricardo Troncoso and Jorge Zanelli for their interesting comments.
XOC wishes to thank the Perimeter Institute for its kind hospitality where part of this research was performed. XOC is supported by a spanish FPU fellowship. He is thankful to the Front of Galician-speaking Scientists for encouragement.
This work is supported in part by MICINN and FEDER (grants FPA2008-01838 and FPA2011-22594), by Xunta de Galicia (Conseller\'{\i}a de Educaci\'on and grant PGIDIT10PXIB206075PR), and by the Spanish Consolider-Ingenio 2010 Programme CPAN (CSD2007-00042).
The Centro de Estudios Cient\'\i ficos (CECS) is funded by the Chilean Government through the Millennium Science Initiative and the Centers of Excellence Base Financing Program of Conicyt, and by the Conicyt grant ÒSouthern Theoretical Physics LaboratoryÓ ACT-91. CECS is also supported by a group of private companies which at present includes Antofagasta Minerals, Arauco, Empresas CMPC, Indura, Naviera Ultragas and Telef\'onica del Sur.
\vskip7mm


\providecommand{\href}[2]{#2}\begingroup\raggedright\endgroup

\begin{thebibliography}{10}

\bibitem{Lovelock1971}
D.~Lovelock, {\it {The Einstein tensor and its generalizations}},  {\em J.
  Math. Phys.} {\bf 12} (1971) 498--501.

\bibitem{Zwiebach1985}
B.~Zwiebach, {\it {Curvature Squared Terms and String Theories}},  {\em Phys.
  Lett.} {\bf B156} (1985) 315.

\bibitem{Maldacena1998}
J.~M. Maldacena, {\it {The large N limit of superconformal field theories and
  supergravity}},  {\em Adv. Theor. Math. Phys.} {\bf 2} (1998) 231--252,
  [\href{http://xxx.lanl.gov/abs/hep-th/9711200}{{\tt hep-th/9711200}}].

\bibitem{Klebanov1998}
I.~R. Klebanov and E.~Witten, {\it {Superconformal field theory on threebranes
  at a Calabi-Yau singularity}},  {\em Nucl. Phys.} {\bf B536} (1998) 199--218,
  [\href{http://xxx.lanl.gov/abs/hep-th/9807080}{{\tt hep-th/9807080}}].

\bibitem{Polyakov1999}
A.~M. Polyakov, {\it {The wall of the cave}},  {\em Int. J. Mod. Phys.} {\bf
  A14} (1999) 645--658, [\href{http://xxx.lanl.gov/abs/hep-th/9809057}{{\tt
  hep-th/9809057}}].

\bibitem{Strominger:1997eq}
A.~Strominger, {\it {Black hole entropy from near horizon microstates}},  {\em
  JHEP} {\bf 9802} (1998) 009,
  [\href{http://xxx.lanl.gov/abs/hep-th/9712251}{{\tt hep-th/9712251}}].

\bibitem{Brown1986b}
J.~D. Brown and M.~Henneaux, {\it {Central charges in the canonical realization
  of asymptotic symmetries: an example from three-dimensional gravity}},  {\em
  Commun. Math. Phys.} {\bf 104} (1986) 207--226.

\bibitem{Witten2007c}
E.~Witten, {\it {Conformal Field Theory In Four And Six Dimensions}},
  \href{http://xxx.lanl.gov/abs/0712.0157}{{\tt arXiv:0712.0157}}.

\bibitem{Boer2009}
J.~de~Boer, M.~Kulaxizi, and A.~Parnachev, {\it {AdS$_7$/CFT$_6$, Gauss-Bonnet
  Gravity, and Viscosity Bound}},  {\em JHEP} {\bf 03} (2010) 087,
  [\href{http://xxx.lanl.gov/abs/0910.5347}{{\tt arXiv:0910.5347}}].

\bibitem{Camanho2010}
X.~O. Camanho and J.~D. Edelstein, {\it {Causality constraints in AdS/CFT from
  conformal collider physics and Gauss-Bonnet gravity}},  {\em JHEP} {\bf 04}
  (2010) 007, [\href{http://xxx.lanl.gov/abs/0911.3160}{{\tt
  arXiv:0911.3160}}].

\bibitem{Buchel2010a}
A.~Buchel, J.~Escobedo, R.~C. Myers, M.~F. Paulos, A.~Sinha, and M.~Smolkin,
  {\it {Holographic GB gravity in arbitrary dimensions}},  {\em JHEP} {\bf 03}
  (2010) 111, [\href{http://xxx.lanl.gov/abs/0911.4257}{{\tt
  arXiv:0911.4257}}].

\bibitem{Camanho2010a}
X.~O. Camanho and J.~D. Edelstein, {\it {Causality in AdS/CFT and Lovelock
  theory}},  {\em JHEP} {\bf 06} (2010) 099,
  [\href{http://xxx.lanl.gov/abs/0912.1944}{{\tt arXiv:0912.1944}}].

\bibitem{Boer2009a}
J.~de~Boer, M.~Kulaxizi, and A.~Parnachev, {\it {Holographic Lovelock gravities
  and black holes}},  {\em JHEP} {\bf 06} (2010) 008,
  [\href{http://xxx.lanl.gov/abs/0912.1877}{{\tt arXiv:0912.1877}}].

\bibitem{Hofman2008}
D.~M. Hofman and J.~Maldacena, {\it {Conformal collider physics: Energy and
  charge correlations}},  {\em JHEP} {\bf 05} (2008) 012,
  [\href{http://xxx.lanl.gov/abs/0803.1467}{{\tt arXiv:0803.1467}}].

\bibitem{Buchel2009a}
A.~Buchel and R.~C. Myers, {\it {Causality of Holographic Hydrodynamics}}, {\em
  JHEP} {\bf 08} (2009) 016, [\href{http://xxx.lanl.gov/abs/0906.2922}{{\tt
  arXiv:0906.2922}}].

\bibitem{Hofman2009}
D.~M. Hofman, {\it {Higher Derivative Gravity, Causality and Positivity of
  Energy in a UV complete QFT}},  {\em Nucl. Phys.} {\bf B823} (2009) 174--194,
  [\href{http://xxx.lanl.gov/abs/0907.1625}{{\tt arXiv:0907.1625}}].

\bibitem{Brigante2008}
M.~Brigante, H.~Liu, R.~C. Myers, S.~Shenker, and S.~Yaida, {\it {Viscosity
  Bound Violation in Higher Derivative Gravity}},  {\em Phys. Rev.} {\bf D77}
  (2008) 126006, [\href{http://xxx.lanl.gov/abs/0712.0805}{{\tt
  arXiv:0712.0805}}].

\bibitem{Kulaxizi2011}
M.~Kulaxizi and A.~Parnachev, {\it {Energy Flux Positivity and Unitarity in
  CFTs}}, {\em Phys. Rev. Lett.} {\bf 106} (2011) 011601,
  [\href{http://xxx.lanl.gov/abs/1007.0553}{{\tt arXiv:1007.0553}}].

\bibitem{Paulos2011}
M.~F. Paulos, {\it {Holographic phase space: $c$-functions and black holes as
  renormalization group flows}}, {\em JHEP} {\bf 05} (2011) 043,
  [\href{http://xxx.lanl.gov/abs/1101.5993}{{\tt arXiv:1101.5993}}].

\bibitem{Maeda2011}
H.~Maeda, S.~Willison, and S.~Ray, {\it {Lovelock black holes with maximally
  symmetric horizons}},  \href{http://xxx.lanl.gov/abs/1103.4184}{{\tt
  arXiv:1103.4184}}.

\bibitem{Zumino1986}
B.~Zumino, {\it {Gravity Theories in More Than Four-Dimensions}},  {\em Phys.
  Rept.} {\bf 137} (1986) 109.

\bibitem{Camanho2012}
X.~O. Camanho and J.~D. Edelstein, to appear.

\bibitem{Aros2001}
R.~Aros, R.~Troncoso, and J.~Zanelli, {\it Black holes with topologically
  nontrivial ads asymptotics},  {\em Phys.Rev. D} {\bf 63} (2001) 084015,
  [\href{http://xxx.lanl.gov/abs/hep-th/0011097}{{\tt hep-th/0011097}}].

\bibitem{Wolf2011}
J.~A. Wolf, {\em Spaces of constant curvature}, AMS Chelsea Publishing Chelsea Publishing, Providence, Rhode Island,
  6th~ed., 2011.

\bibitem{Zegers2005}
R.~Zegers, {\it Birkhoff's theorem in lovelock gravity},  {\em J. Math. Phys.}
  {\bf 46} (2005) 072502,
  [\href{http://xxx.lanl.gov/abs/gr-qc/0505016}{{\tt gr-qc/0505016}}].

\bibitem{Charmousis2002a}
C.~Charmousis and J.-F. Dufaux, {\it {General Gauss-Bonnet brane cosmology}},
  {\em Class. Quant. Grav.} {\bf 19} (2002) 4671--4682,
  [\href{http://xxx.lanl.gov/abs/hep-th/0202107}{{\tt hep-th/0202107}}].

\bibitem{Kastor2010}
D.~Kastor, S.~Ray, and J.~Traschen, {\it {Smarr formula and an extended first
  law for Lovelock gravity}},  {\em Class. Quant. Grav.} {\bf 27} (2010) 235014
  [\href{http://xxx.lanl.gov/abs/1005.5053}{{\tt arXiv:1005.5053}}].

\bibitem{Kastor2011}
D.~Kastor, S.~Ray, and J.~Traschen, {\it {Mass and free energy of Lovelock black
  holes}},  {\em Class. Quant. Grav.} {\bf 28} (2011) 195022
  [\href{http://xxx.lanl.gov/abs/1106.2764}{{\tt arXiv:1106.2764}}].

\bibitem{Boulware1985a}
D.~G. Boulware and S.~Deser, {\it {String Generated Gravity Models}},  {\em
  Phys. Rev. Lett.} {\bf 55} (1985) 2656.

\bibitem{Wheeler1986}
J.~T. Wheeler, {\it {Symmetric Solutions to the Gauss-Bonnet Extended Einstein
  Equations}},  {\em Nucl. Phys.} {\bf B268} (1986) 737.

\bibitem{Wheeler1986a}
J.~T. Wheeler, {\it {Symmetric solutions to the maximally Gauss-Bonnet extended
  Einstein equations}},  {\em Nucl. Phys.} {\bf B273} (1986) 732.

\bibitem{Cai1999}
R.-G. Cai and K.-S. Soh, {\it {Topological black holes in the dimensionally
  continued gravity}},  {\em Phys. Rev.} {\bf D59} (1999) 044013,
  [\href{http://xxx.lanl.gov/abs/gr-qc/9808067}{{\tt gr-qc/9808067}}].

\bibitem{Myers2010b}
R.~C. Myers and B.~Robinson, {\it {Black Holes in Quasi-topological Gravity}},
  {\em JHEP} {\bf 08} (2010) 067,
  [\href{http://xxx.lanl.gov/abs/1003.5357}{{\tt arXiv:1003.5357}}].

\bibitem{Marden1949}
M.~Marden, {\it {Geometry of polynomials}},  {\em Mathematical Surveys and
  Monographs} {\bf 3} (1949). Providence, USA: Am. Math. Soc. (1949) 243 p.

\bibitem{Camanho-u}
X.~O. Camanho, unpublished.

\bibitem{Charmousis2008a}
C.~Charmousis and A.~Padilla, {\it {The Instability of Vacua in Gauss-Bonnet
  Gravity}},  {\em JHEP} {\bf 12} (2008) 038,
  [\href{http://xxx.lanl.gov/abs/0807.2864}{{\tt arXiv:0807.2864}}].

\bibitem{Mann1997a}
R.~B. Mann, {\it {Black holes of negative mass}},  {\em Class. Quant. Grav.}
  {\bf 14} (1997) 2927--2930,
  [\href{http://xxx.lanl.gov/abs/gr-qc/9705007}{{\tt gr-qc/9705007}}].

\bibitem{Horowitz1998}
G.~T. Horowitz and R.~C. Myers, {\it {The AdS/CFT Correspondence and a New
  Positive Energy Conjecture for General Relativity}},  {\em Phys. Rev.} {\bf
  D59} (1998) 026005, [\href{http://xxx.lanl.gov/abs/hep-th/9808079}{{\tt
  hep-th/9808079}}].

\bibitem{Cai2004}
R.-G. Cai, {\it {A note on thermodynamics of black holes in Lovelock gravity}},
   {\em Phys. Lett.} {\bf B582} (2004) 237--242,
  [\href{http://xxx.lanl.gov/abs/hep-th/0311240}{{\tt hep-th/0311240}}].

\bibitem{Witten1998}
E.~Witten, {\it {Anti-de Sitter space and holography}},  {\em Adv. Theor. Math.
  Phys.} {\bf 2} (1998) 253--291,
  [\href{http://xxx.lanl.gov/abs/hep-th/9802150}{{\tt hep-th/9802150}}].

\bibitem{Witten1998a}
E.~Witten, {\it {Anti-de Sitter space, thermal phase transition, and
  confinement in gauge theories}},  {\em Adv. Theor. Math. Phys.} {\bf 2}
  (1998) 505--532, [\href{http://xxx.lanl.gov/abs/hep-th/9803131}{{\tt
  hep-th/9803131}}].

\bibitem{Nojiri2001j}
S.~Nojiri and S.~D. Odintsov, {\it {Anti-de Sitter black hole thermodynamics in
  higher derivative gravity and new confining-deconfining phases in dual CFT}},
   {\em Phys. Lett.} {\bf B521} (2001) 87--95,
  [\href{http://xxx.lanl.gov/abs/hep-th/0109122}{{\tt hep-th/0109122}}].

\bibitem{Cho2002a}
Y.~M. Cho and I.~P. Neupane, {\it {Anti-de Sitter black holes, thermal phase
  transition and holography in higher curvature gravity}},  {\em Phys. Rev.}
  {\bf D66} (2002) 024044, [\href{http://xxx.lanl.gov/abs/hep-th/0202140}{{\tt
  hep-th/0202140}}].

\bibitem{Myers1988}
R.~C. Myers and J.~Z. Simon, {\it {Black Hole Thermodynamics in Lovelock
  Gravity}},  {\em Phys. Rev.} {\bf D38} (1988) 2434--2444.

\bibitem{Nojiri2001n}
S.~Nojiri and S.~D. Odintsov, {\it {The de Sitter/anti-de Sitter black holes
  phase transition?}},  \href{http://xxx.lanl.gov/abs/gr-qc/0112066}{{\tt
  gr-qc/0112066}}.

\bibitem{Cvetic2002}
M.~Cvetic, S.~Nojiri, and S.~D. Odintsov, {\it {Black hole thermodynamics and
  negative entropy in de Sitter and anti-de Sitter Einstein-Gauss-Bonnet
  gravity}},  {\em Nucl. Phys.} {\bf B628} (2002) 295--330,
  [\href{http://xxx.lanl.gov/abs/hep-th/0112045}{{\tt hep-th/0112045}}].

\bibitem{Neupane2009a}
I.~P. Neupane and N.~Dadhich, {\it {Entropy bound and causality violation in
  higher curvature gravity}},  {\em Class. Quant. Grav.} {\bf 26} (2009)
  015013.

\bibitem{Neupane2009f}
I.~P. Neupane, {\it {Black Holes, Entropy Bound and Causality Violation}},
  {\em Int. J. Mod. Phys.} {\bf A24} (2009) 3584--3591,
  [\href{http://xxx.lanl.gov/abs/0904.4805}{{\tt arXiv:0904.4805}}].

\bibitem{Emparan1999}
R.~Emparan, C.~V. Johnson, and R.~C. Myers, {\it {Surface terms as counterterms
  in the AdS/CFT correspondence}},  {\em Phys. Rev.} {\bf D60} (1999) 104001,
  [\href{http://xxx.lanl.gov/abs/hep-th/9903238}{{\tt hep-th/9903238}}].

\bibitem{Jacobson1993}
T.~Jacobson and R.~C. Myers, {\it {Black hole entropy and higher curvature
  interactions}},  {\em Phys. Rev. Lett.} {\bf 70} (1993) 3684--3687,
  [\href{http://xxx.lanl.gov/abs/hep-th/9305016}{{\tt hep-th/9305016}}].

\bibitem{Neupane2004}
I.~P. Neupane, {\it {Thermodynamic and gravitational instability on hyperbolic
  spaces}},  {\em Phys. Rev.} {\bf D69} (2004) 084011,
  [\href{http://xxx.lanl.gov/abs/hep-th/0302132}{{\tt hep-th/0302132}}].

\bibitem{Gibbons1977a}
G.~W. Gibbons and S.~W. Hawking, {\it {Cosmological event horizons,
  thermodynamics, and particle creation}},  {\em Phys. Rev.} {\bf D15} (1977)
  2738--2751.

\bibitem{Vanzo1997a}
L.~Vanzo, {\it {Black holes with unusual topology}},  {\em Phys. Rev. D} {\bf
  D56} (1997) 6475--6483.

\bibitem{Birmingham1999}
D.~Birmingham, {\it {Topological black holes in anti-de Sitter space}},  {\em
  Class. Quant. Grav.} {\bf 16} (1999) 1197--1205,
  [\href{http://xxx.lanl.gov/abs/hep-th/9808032}{{\tt hep-th/9808032}}].

\bibitem{Ginsparg1983}
P.~H. Ginsparg and M.~J. Perry, {\it {Semiclassical Perdurance of de Sitter
  Space}},  {\em Nucl. Phys.} {\bf B222} (1983) 245.

\bibitem{Torii2005}
T.~Torii and H.~Maeda, {\it {Spacetime structure of static solutions in
  Gauss-Bonnet gravity: Neutral case}},  {\em Phys. Rev.} {\bf D71} (2005)
  124002, [\href{http://xxx.lanl.gov/abs/hep-th/0504127}{{\tt
  hep-th/0504127}}].

\bibitem{Torii2005a}
T.~Torii and H.~Maeda, {\it {Spacetime structure of static solutions in
  Gauss-Bonnet gravity: Charged case}},  {\em Phys. Rev.} {\bf D72} (2005)
  064007, [\href{http://xxx.lanl.gov/abs/hep-th/0504141}{{\tt
  hep-th/0504141}}].

\bibitem{Cai2002}
R.-G. Cai, {\it {Gauss-Bonnet black holes in AdS spaces}},  {\em Phys. Rev.}
  {\bf D65} (2002) 084014, [\href{http://xxx.lanl.gov/abs/hep-th/0109133}{{\tt
  hep-th/0109133}}].

\bibitem{Matzner1979}
R.~A. Matzner, N.~Zamorano, and V.~D. Sandberg, {\it {Instability of the Cauchy
  horizon of Reissner-Nordstrom black holes}},  {\em Phys. Rev.} {\bf D19}
  (1979) 2821--2826.

\bibitem{Poisson1990}
E.~Poisson and W.~Israel, {\it {Internal structure of black holes}},  {\em
  Phys. Rev.} {\bf D41} (1990) 1796--1809.

\bibitem{Brady1995}
P.~R. Brady and J.~D. Smith, {\it {Black hole singularities: A Numerical
  approach}},  {\em Phys. Rev. Lett.} {\bf 75} (1995) 1256--1259,
  [\href{http://xxx.lanl.gov/abs/gr-qc/9506067}{{\tt gr-qc/9506067}}].

\bibitem{Dafermos2003}
M.~Dafermos, {\it {The interior of charged black holes and the problem of
  uniqueness in general relativity}},
  \href{http://xxx.lanl.gov/abs/gr-qc/0307013}{{\tt gr-qc/0307013}}.

\bibitem{Torii2006}
T.~Torii, {\it {Black holes in higher curvature theory and third law of
  thermodynamics}},  {\em J. Phys. Conf. Ser.} {\bf 31} (2006) 175--176.

\bibitem{Anderson1995a}
P.~R. Anderson, W.~A. Hiscock, and D.~J. Loranz, {\it {Semiclassical stability
  of the extreme Reissner-Nordstrom black hole}},  {\em Phys. Rev. Lett.} {\bf
  74} (1995) 4365--4368, [\href{http://xxx.lanl.gov/abs/gr-qc/9504019}{{\tt
  gr-qc/9504019}}].

\bibitem{Marolf2010}
D.~Marolf, {\it {The dangers of extremes}},  {\em Gen. Rel. Grav.} {\bf 42}
  (2010) 2337--2343, [\href{http://xxx.lanl.gov/abs/1005.2999}{{\tt
  arXiv:1005.2999}}].

\bibitem{Camanho2012b}
X.~O.~Camanho, J.~D.~Edelstein, G.~Giribet, and A.~Gomberoff, {\it {A new type
of phase transition in gravitational theories}},
\href{http://xxx.lanl.gov/abs/1204.6737}{{\tt arXiv:1204.6737}}.

\bibitem{Dotti2005b}
G.~Dotti and R.~J. Gleiser, {\it {Linear stability of Einstein-gauss-bonnet
  static spacetimes. part. I: Tensor perturbations}},  {\em Phys. Rev.} {\bf
  D72} (2005) 044018, [\href{http://xxx.lanl.gov/abs/gr-qc/0503117}{{\tt
  gr-qc/0503117}}].

\bibitem{Gleiser2005b}
R.~J. Gleiser and G.~Dotti, {\it {Linear stability of Einstein-Gauss-Bonnet
  static spacetimes. II: Vector and scalar perturbations}},  {\em Phys. Rev.}
  {\bf D72} (2005) 124002, [\href{http://xxx.lanl.gov/abs/gr-qc/0510069}{{\tt
  gr-qc/0510069}}].

\bibitem{Beroiz2007}
M.~Beroiz, G.~Dotti, and R.~J. Gleiser, {\it {Gravitational instability of
  static spherically symmetric Einstein-Gauss-Bonnet black holes in five and
  six dimensions}},  {\em Phys. Rev.} {\bf D76} (2007) 024012,
  [\href{http://xxx.lanl.gov/abs/hep-th/0703074}{{\tt hep-th/0703074}}].

\bibitem{Takahashi2009h}
T.~Takahashi and J.~Soda, {\it {Stability of Lovelock Black Holes under Tensor
  Perturbations}},  {\em Phys. Rev.} {\bf D79} (2009) 104025,
  [\href{http://xxx.lanl.gov/abs/0902.2921}{{\tt arXiv:0902.2921}}].

\bibitem{Takahashi2009d}
T.~Takahashi and J.~Soda, {\it {Instability of Small Lovelock Black Holes in
  Even- dimensions}},  {\em Phys. Rev.} {\bf D80} (2009) 104021,
  [\href{http://xxx.lanl.gov/abs/0907.0556}{{\tt arXiv:0907.0556}}].

\bibitem{Takahashi2010e}
T.~Takahashi and J.~Soda, {\it {Catastrophic Instability of Small Lovelock
  Black Holes}},  {\em Prog. Theor. Phys.} {\bf 124} (2010) 711--729,
  [\href{http://xxx.lanl.gov/abs/1008.1618}{{\tt arXiv:1008.1618}}].

\bibitem{Takahashi2010d}
T.~Takahashi and J.~Soda, {\it {Master Equations for Gravitational
  Perturbations of Static Lovelock Black Holes in Higher Dimensions}},  {\em
  Prog. Theor. Phys.} {\bf 124} (2010) 911--924,
  [\href{http://xxx.lanl.gov/abs/1008.1385}{{\tt arXiv:1008.1385}}].

\bibitem{Camanho2010d}
X.~O. Camanho, J.~D. Edelstein, and M.~F. Paulos, {\it {Lovelock theories,
  holography and the fate of the viscosity bound}},
  \href{http://xxx.lanl.gov/abs/1010.1682}{{\tt arXiv:1010.1682}}.

\bibitem{Crisostomo2000}
J.~Crisostomo, R.~Troncoso, and J.~Zanelli, {\it {Black hole scan}},  {\em
  Phys. Rev.} {\bf D62} (2000) 084013,
  [\href{http://xxx.lanl.gov/abs/hep-th/0003271}{{\tt hep-th/0003271}}].

\bibitem{Cai2004a}
R.-G. Cai and Q.~Guo, {\it {Gauss-Bonnet black holes in dS spaces}},  {\em
  Phys. Rev.} {\bf D69} (2004) 104025,
  [\href{http://xxx.lanl.gov/abs/hep-th/0311020}{{\tt hep-th/0311020}}].

\bibitem{Hawking1995}
S.~W. Hawking, G.~T. Horowitz, and S.~F. Ross, {\it {Entropy, Area, and black
  hole pairs}},  {\em Phys. Rev.} {\bf D51} (1995) 4302--4314,
  [\href{http://xxx.lanl.gov/abs/gr-qc/9409013}{{\tt gr-qc/9409013}}].

\bibitem{Gravanis2010a}
E.~Gravanis, {\it {Shock waves and Birkhoff's theorem in Lovelock gravity}},
  {\em Phys. Rev.} {\bf D82} (2010) 104024,
  [\href{http://xxx.lanl.gov/abs/1008.3583}{{\tt arXiv:1008.3583}}].

\bibitem{Camanho2012c}
X.~O.~Camanho, J.~D.~Edelstein, G.~Giribet, and A.~Gomberoff, to appear.

\bibitem{Deser1982a}
S.~Deser, R.~Jackiw, and S.~Templeton, {\it {Three-Dimensional Massive Gauge
  Theories}},  {\em Phys. Rev. Lett.} {\bf 48} (1982) 975--978.

\bibitem{Oliva2010}
J.~Oliva and S.~Ray, {\it {A new cubic theory of gravity in five dimensions:
  Black hole, Birkhoff's theorem and C-function}},  {\em Class. Quant. Grav.}
  {\bf 27} (2010) 225002, [\href{http://xxx.lanl.gov/abs/1003.4773}{{\tt
  arXiv:1003.4773}}].

\bibitem{Oliva2010a}
J.~Oliva and S.~Ray, {\it {Classification of Six Derivative Lagrangians of
  Gravity and Static Spherically Symmetric Solutions}},  {\em Phys. Rev.} {\bf
  D82} (2010) 124030, [\href{http://xxx.lanl.gov/abs/1004.0737}{{\tt
  arXiv:1004.0737}}].

\bibitem{Myers2010d}
R.~C. Myers, M.~F. Paulos, and A.~Sinha, {\it {Holographic studies of
  quasi-topological gravity}},  {\em JHEP} {\bf 08} (2010) 035,
  [\href{http://xxx.lanl.gov/abs/1004.2055}{{\tt arXiv:1004.2055}}].

\end{thebibliography}
\end{document}